# Lattice Boltzmann scheme for hydrodynamic equation of phonon transport

Yangyu Guo, Moran Wang[†]

*Department of Engineering Mechanics and CNMM, Tsinghua University, Beijing 100084, China*

## Abstract

In this work, a lattice Boltzmann scheme is developed for numerical solution of the phonon Boltzmann equation under Callaway's dual relaxation model in the hydrodynamic limit. Through a Chapman-Enskog expansion to the lattice Boltzmann equation with the resistive scattering term as an equivalent source term, we recover a phonon hydrodynamic equation which is reduced to the Guyer-Krumhansl heat transport equation and the Fourier's law in the limit of dominant normal scattering and dominant resistive scattering respectively. Several cases of heat transport from diffusive regime to hydrodynamic regime are modeled extensively by the present numerical scheme, which produces results in good agreement with the benchmark solutions. Two well-known phonon hydrodynamic phenomena including the phonon Poiseuille flow and second sound propagation are well captured by the lattice Boltzmann scheme. This work will promote the numerical modeling and deeper understanding of the non-Fourier heat transport induced by phonon normal scattering.

*Keywords*: phonon transport; lattice Boltzmann method; Callaway model; phonon hydrodynamics; non-Fourier conduction

---

[†] Corresponding author; Tel: 86-10-627-87498; Email: mrwang@tsinghua.edu.cn

# 1. Introduction

The classical Fourier's law has been widely applied in modeling heat transport in conventional systems and processes [1, 2]. In dielectric or semiconductor materials, heat transport is mainly mediated by phonons. The Fourier's law is valid in diffusive regime where the intrinsic phonon resistive scattering dominates [3]. With the rapid development of micro- and nano-electronics in recent years, the Fourier's law fails in extremely-small and ultra-fast heat transport [4-6] due to both spatial and temporal strong nonequilibrium effects [7, 8]. Another situation where the Fourier's law fails is in the hydrodynamic transport regime where the momentum-conserving phonon normal scattering dominates [9, 10]. This situation is often relevant at extremely low temperature in most three-dimensional materials, and has drawn increasing attention in recent years because of the strong evidence of hydrodynamic phonon transport even near room temperature in low-dimensional materials like graphene [11-14] and carbon nanotubes [15].

The theoretical modeling of hydrodynamic phonon transport has an appreciably long history. The classical Guyer-Krumhansl (G-K) heat transport equation was derived through a solution of the linearized phonon Boltzmann equation during the exploration of second sound in the middle years of last century [9, 10]. A classical simplification of the complex scattering term in the phonon Boltzmann equation is the Callaway's dual relaxation model [16], which has been widely used for studying phonon hydrodynamic phenomena in three-dimensional dielectric crystals [8, 17-19] as well as in low-dimensional materials [12-15]. As the Callaway model considers the normal and resistive phonon scattering separately, it is capable of describing heat transport in a broad range from the hydrodynamic regime to the diffusive regime.

The numerical solution of the phonon Boltzmann equation under Callaway's dual relaxation model remains, however, in its infancy. A discrete ordinate method (DOM) has been recently developed for the Callaway's kinetic model to study heat transport in two-dimensional nanomaterials [13]. Yet the DOM solution of Boltzmann equation is known to be inefficient within the near-continuum regime as the spatial step and temporal step are required to be smaller than the mean free path and relaxation time [20, 21]. In this regime, the hydrodynamic equations of phonon Boltzmann equation provide a more practical theoretical description. In contrast to traditional numerical methods for partial differential equations, the lattice Boltzmann scheme is a more efficient one for hydrodynamic equations through an evolution of discrete Boltzmann equation with a linear advection term and an algebraic collision term [22-24]. A lattice Boltzmann scheme has been presented through a Chapman-Enskog expansion to the phonon lattice Boltzmann equation under the Callaway model [25] suggested by an earlier work on lattice Boltzmann model for phonon hydrodynamics with only normal scattering considered [26]. An interfacial boundary collision scheme was later devised for this lattice scheme to simulate the interfacial phonon transport between dissimilar materials [27, 28]. These studies [25, 27] represent crucial progress in the development of lattice Boltzmann scheme for the Callaway model, yet with still aspects to be improved [8]: the phonon hydrodynamic equation has not been exactly recovered, such that the relation between the mesoscopic numerical parameters and macroscopic phonon properties shall be further clarified. The similar issues have been recently resolved in the lattice Boltzmann scheme for the phonon Boltzmann equation under single mode relaxation time approximation [29]. However, a lattice Boltzmann scheme for the phonon Boltzmann equation under Callaway's dual relaxation model in the hydrodynamic limit remains to be fully established.

The main aim of the present work is to develop a lattice Boltzmann scheme for numerical solution of phonon Boltzmann equation under Callaway's dual relaxation model within the near-continuum regime. Based on the lattice Boltzmann model, we also reveal some special physical features of non-Fourier heat conduction induced by the collective effect of phonon normal scattering. The remainder of this article is organized as below: in Section 2, the mathematical and numerical methodology is provided in details; in Section 3, a robust validation is demonstrated of the lattice Boltzmann scheme through both steady-state and transient heat transport from diffusive regime to hydrodynamic regime; concluding remarks are finally made in Section 4.

## 2. Mathematical and Numerical Methodology

In this section, we firstly present the hydrodynamic equation to the phonon Boltzmann equation under the Callaway model in Section 2.1. The lattice Boltzmann scheme for the Callaway model in the hydrodynamic limit is then developed in Section 2.2. Through a Chapman-Enskog expansion to the lattice Boltzmann equation, we recover the phonon hydrodynamic equation and obtain the quantitative relation between numerical parameters and phonon properties in Section 2.3. Finally the macroscopic variable calculation and boundary treatment are provided in Section 2.4 and 2.5 respectively.

### 2.1. Phonon hydrodynamic equations

The evolution of lattice Boltzmann equation corresponds to the numerical solution of the hydrodynamic equation of the continuous Boltzmann equation [22-24]. In the lattice scheme for the phonon Boltzmann equation under single mode relaxation time approximation, the hydrodynamic equation is the Fourier's law [29]. Here we provide a derivation of hydrodynamic equation to the phonon Boltzmann equation under Callaway's dual relaxation model by a perturbation method [30].

The phonon Boltzmann equation under Callaway's dual relaxation model is [16]:

$$\frac{\partial f}{\partial t} + \mathbf{v}_g \cdot \nabla_\mathbf{x} f = \frac{f_N^{eq} - f}{\tau_N} + \frac{f_R^{eq} - f}{\tau_R}, \tag{1}$$

where $f \equiv f(t, \mathbf{x}, \mathbf{k})$ is the phonon distribution function, with $f(t, \mathbf{x}, \mathbf{k}) d\mathbf{x} d\mathbf{k}$ denoting the probabilistic number of phonons found within the spatial interval ($\mathbf{x}$, $\mathbf{x}+d\mathbf{x}$) and wave vector interval ($\mathbf{k}$, $\mathbf{k}+d\mathbf{k}$) at a specific time $t$. The phonon group velocity is determined from the phonon dispersion relation as: $\mathbf{v}_g = \nabla_\mathbf{k} \omega(\mathbf{k})$. The equilibrium distribution functions for the normal scattering and the resistive scattering (usually including the umklapp phonon-phonon scattering, phonon-isotope scattering, phonon-impurity scattering, *etc.*) are the displaced Planck distribution and the Planck distribution separately:

$$f_{\mathrm{N}}^{\mathrm{eq}} = \frac{1}{\exp\left[\left(\hbar\omega - \hbar\mathbf{k}\cdot\mathbf{u}\right)/k_{\mathrm{B}}T\right]-1}, \tag{2}$$

$$f_{\mathrm{R}}^{\mathrm{eq}} = \frac{1}{\exp\left(\hbar\omega/k_{\mathrm{B}}T\right)-1}, \tag{3}$$

where $\hbar$ and $k_{\mathrm{B}}$ are the reduced Planck constant and Boltzmann constant separately. The temperature $T$ and phonon drift velocity $\mathbf{u}$ can be determined by the energy and momentum conservation relations: $\int \hbar\omega f d\mathbf{k} = \int \hbar\omega f_{\mathrm{R}}^{\mathrm{eq}} d\mathbf{k}$ and $\int \hbar\mathbf{k} f d\mathbf{k} = \int \hbar\mathbf{k} f_{\mathrm{N}}^{\mathrm{eq}} d\mathbf{k}$. The phonon relaxation times for normal scattering ($\tau_{\mathrm{N}}$) and resistive scattering ($\tau_{\mathrm{R}}$) are usually dependent on the phonon frequency and lattice temperature [31]. As a first step, the isotropic gray Debye model is assumed for the derivation of hydrodynamic equation and lattice Boltzmann scheme. In other words, we consider a linear phonon dispersion $\omega = v_{\mathrm{g}}k$ with constant group velocity, and constant effective phonon relaxation times around an average system temperature. When the heat transport is driven under a relatively small temperature difference, the displaced Planck distribution is often linearized as [16]:

$$f_{\mathrm{N}}^{\mathrm{eq}} = f_{\mathrm{R}}^{\mathrm{eq}} + T\frac{\partial f_{\mathrm{R}}^{\mathrm{eq}}}{\partial T}\frac{\mathbf{k}\cdot\mathbf{u}}{\omega}. \tag{4}$$

The main idea of the perturbation method is an asymptotic expansion to the phonon Boltzmann equation (1) around a four-moment non-equilibrium distribution obtained by maximum entropy principle [17, 30]:

$$f_4 = f_{\mathrm{R}}^{\mathrm{eq}} + \frac{3}{C_V v_{\mathrm{g}}^2}\frac{\partial f_{\mathrm{R}}^{\mathrm{eq}}}{\partial T}q_{\alpha}v_{\mathrm{g}\alpha}, \tag{5}$$

where the subscript '4' denotes its dependence on the four basic field variables (temperature and three components of heat flux). The Einstein summation convention is adopted hereafter, with $\alpha$ denoting the components ($x$, $y$, $z$) of cartesian coordinate. As this work is focused on the development of lattice Boltzmann scheme, here we only provide the final results of phonon hydrodynamic equations including an energy balance equation and a heat transport equation:

$$C_V \frac{\partial T}{\partial t} + \nabla \cdot \mathbf{q} = 0 \,, \tag{6}$$

$$\tau_{\mathrm{R}} \frac{\partial \mathbf{q}}{\partial t} + \mathbf{q} = -\frac{1}{3} C_V v_g^2 \tau_{\mathrm{R}} \nabla T + \frac{1}{5} \tau_{\mathrm{C}} \tau_{\mathrm{R}} v_g^2 \left[ \nabla^2 \mathbf{q} + \frac{1}{3} \nabla (\nabla \cdot \mathbf{q}) \right] , \tag{7}$$

where the overall phonon relaxation time is defined as: $1/\tau_{\mathrm{C}} = 1/\tau_{\mathrm{N}} + 1/\tau_{\mathrm{R}}$. The details of the derivation of phonon hydrodynamic equations are given in Appendix A. Note two similar field equations as Eq. (6) and Eq. (7) have been obtained in Ref. [32] in the framework of nine-moment phonon hydrodynamic equations. However, there is a subtle difference that they were derived in the limit of small overall Knudsen number $Kn_{\mathrm{C}}$ but large resistive Knudsen number $Kn_{\mathrm{R}}$, i.e. $Kn_{\mathrm{C}} \approx Kn_{\mathrm{N}} << Kn_{\mathrm{R}}$ or $\tau_{\mathrm{C}} \approx \tau_{\mathrm{N}} << \tau_{\mathrm{R}}$ (c.f. the derivation of Eq. (52) in Ref. [32], or a more detailed derivation in the limit of small $\tau_{\mathrm{N}}$ in Section 5.2.2 of Ref. [17]). In other words, they concern the case of dominant normal scattering over resistive scattering. Actually, the two field equations in Ref. [32] were asserted therein to be equivalent to the original G-K heat transport equation (except a factor difference) which was derived in the limit of dominant normal scattering [9, 33]. In this work, we are concerning the hydrodynamic equation for more general situation including the entire transition from dominant normal scattering to dominant resistive scattering. In addition, we provide a quantitative validation of Eq. (6) and Eq. (7) through a comparison to the direct DOM solution of phonon Boltzmann equation under the Callaway model [13] as given in Appendix A.

In the limit of dominant normal scattering over resistive scattering ($1/\tau_{\mathrm{N}} >> 1/\tau_{\mathrm{R}}$), the overall phonon relaxation time becomes $1/\tau_{\mathrm{C}} \approx 1/\tau_{\mathrm{N}}$. Then Eq. (7) is reduced to the classical G-K heat transport equation [8, 10, 32]:

$$\tau_{\mathrm{R}} \frac{\partial \mathbf{q}}{\partial t} + \mathbf{q} = -\frac{1}{3} C_V v_g^2 \tau_{\mathrm{R}} \nabla T + \frac{1}{5} \tau_{\mathrm{N}} \tau_{\mathrm{R}} v_g^2 \left[ \nabla^2 \mathbf{q} + \frac{1}{3} \nabla (\nabla \cdot \mathbf{q}) \right] . \tag{8}$$

A slightly different coefficient '2' in the nonlocal term of heat flux in the original G-K heat transport equation [10] from a coefficient '1/3' in Eq. (8) has been attributed to their approximate treatment of the inverse of the normal process collision matrix [33]. In the limit of dominant resistive scattering over the normal scattering ($1/\tau_{\mathrm{R}} >> 1/\tau_{\mathrm{N}}$), the overall phonon relaxation time becomes $1/\tau_{\mathrm{C}} \approx 1/\tau_{\mathrm{R}}$. Eq. (7) is then reduced to the recent phonon hydrodynamic equation for nanoscale heat transport at ordinary temperatures [30, 34]:

$$\tau_R \frac{\partial \mathbf{q}}{\partial t} + \mathbf{q} = -\frac{1}{3} C_V v_g^2 \tau_R \nabla T + \frac{1}{5} \Lambda_R^2 \left[ \nabla^2 \mathbf{q} + \frac{1}{3} \nabla (\nabla \cdot \mathbf{q}) \right], \tag{9}$$

where the phonon mean free path for resistive scattering is computed as: $\Lambda_R = v_g \tau_R$. Eq. (9) will be further reduced to the Fourier's law in the diffusive limit where both the memory and nonlocal effects due to the finite relaxation time and mean free path are negligibly weak. Although Eq. (9) has the same mathematical form as Eq. (8), the nonlocal terms of heat flux hold very different physical meaning [30]: (*i*) the transfer of phonon quasi-momentum through the momentum-conserving normal scattering in Eq. (8); (*ii*) representative of spatial non-equilibrium effects from phonon-boundary scattering or large spatial thermal variation in Eq. (9).

Based on the general principle of kinetic theory [35, 36], the temperature and heat flux boundary conditions for the phonon hydrodynamic equation (7) are developed as:

$$T_s - T = -\frac{2}{3} \Lambda_R \left( \frac{\partial T}{\partial n} \right)_s, \tag{10}$$

$$\mathbf{q}_s = \frac{8}{15} \Lambda_C \left( \frac{\partial \mathbf{q}}{\partial n} \right)_s, \tag{11}$$

with the overall phonon mean free path defined as: $\Lambda_C = v_g \tau_C$, and the subscript 's' denoting the boundary surface. The detailed derivation of Eq. (10) and Eq. (11) is given in Appendix A. In the limit of dominant normal scattering, Eq. (11) will be reduced to the heat flux slip boundary condition for G-K heat transport equation (8) [32]. In the limit of dominant resistive scattering, Eq. (11) will be reduced to the heat flux tangential retardant boundary condition for the phonon hydrodynamic equation (9) [30]. Therefore, the present phonon hydrodynamic equation (7) together with the boundary conditions Eq. (10) and Eq. (11) is able to describe the entire transition of heat transport from the hydrodynamic regime to the diffusive regime (quantitative demonstration shown in Appendix A).

## 2.2. Phonon lattice Boltzmann equation

In this sub-section, a lattice Boltzmann scheme will be developed for the hydrodynamic equations of phonon Boltzmann equation (1) under the Callaway model, which is rewritten into the following form:

$$\frac{\partial f}{\partial t} + v_{g\alpha}\frac{\partial f}{\partial x_\alpha} = \frac{f_4 - f}{\tau_C} - \frac{3}{\tau_R C_V v_g^2}\frac{\partial f_R^{eq}}{\partial T}v_{g\alpha}q_\alpha , \tag{12}$$

where the displaced Planck distribution function Eq. (4) has become identical to the non-equilibrium distribution Eq. (5) with the help of the relation $\mathbf{q}=C_V T\mathbf{u}/3$ from the quasi-momentum conservation principle for phonon normal scattering process [8]. Therefore, the phonon Boltzmann equation under Callaway's dual relaxation model is reformulated into an equation under an effective single relaxation model with an equivalent source term. The effective single relaxation scattering term will obey both the energy and quasi-momentum conservation principles. The equivalent source term can be understood as the transport resistance caused by the momentum-destroying phonon resistive scattering.

Then the following lattice Boltzmann equation is devised for Eq. (12) in form of phonon energy density distribution:

$$e_i\left(\mathbf{x}+\mathbf{c}_i\Delta t, t+\Delta t\right) - e_i\left(\mathbf{x}, t\right) = \frac{1}{\tau}\left[e_i^{eq}\left(\mathbf{x}, t\right) - e_i\left(\mathbf{x}, t\right)\right] + \Delta t S_i\left(\mathbf{x}, t\right) , \tag{13}$$

where $\tau$ is a non-dimensional relaxation time related to the overall relaxation time $\tau_C$ and the time step $\Delta t$, with the relation to be determined later. The subscript '$i$' ($i$=1,2, …, 8) represents the index of different lattice directions in the typical D2Q8 lattice structure as shown in Figure 1. The lattice velocities in two-dimension are expressed as:

$$\mathbf{c}_i = \begin{cases} (\pm 1, 0)c, (0, \pm 1)c, & i = 1, 2, 3, 4 \\ (\pm 1, \pm 1)c, & i = 5, 6, 7, 8 \end{cases} , \tag{14}$$

where $c$ is the lattice speed related to phonon group speed $v_g$. The central-point component in the usual D2Q9 lattice structure is not considered because of the special feature of phonon dynamics [29]. The present theoretical derivation can be extended to the one-dimensional (D1Q2) and three-dimensional (D3Q14) lattice structures [29] in a straightforward way, which is not dealt with here due to the length of article.

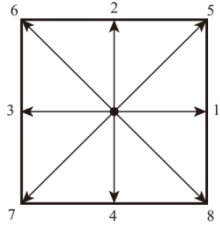

Figure 1. Schematic of D2Q8 lattice structure in the lattice Boltzmann scheme for phonon hydrodynamic equation

The form of discrete equilibrium energy density distribution can be inferred from the four-moment nonequilibrium distribution $f_4$ in Eq. (5) as:

$$e_i^{\text{eq}}(\mathbf{x},t) = \omega_i C_V T(\mathbf{x},t) + \frac{5a_i c_{i\alpha} q_\alpha(\mathbf{x},t)}{3c^2}. \qquad (15)$$

The source term in the lattice Boltzmann equation (13) is defined as:

$$S_i(\mathbf{x},t) = -\left(1 - \frac{1}{2\tau}\right)\frac{5a_i c_{i\alpha} q_\alpha(\mathbf{x},t)}{3\tau_R c^2}, \qquad (16)$$

where the coefficient $(1-1/2\tau)$ is introduced to eliminate the discrete error in the lattice Boltzmann scheme with source term [37]. The weight coefficients in Eq. (15) and Eq. (16) are given as below:

$$\omega_i = \begin{cases} \dfrac{2}{9}, & i=1,2,3,4 \\ \dfrac{1}{36}, & i=5,6,7,8 \end{cases}, \qquad (17)$$

$$a_i = \begin{cases} \dfrac{1}{5}, & i=1,2,3,4 \\ \dfrac{1}{20}, & i=5,6,7,8 \end{cases}. \qquad (18)$$

In principle, the weight coefficients in Eq. (17) and Eq. (18) are unknown at this step, and will be determined after recovering the phonon hydrodynamic equation through Chapman-Enskog expansion in the next sub-section. For a more clear interpretation of the physics of phonon transport, we present the final results firstly and verify them in recovering the hydrodynamic equations. In contrast to the same set of weight coefficients in the discrete equilibrium distribution and the forcing terms in classical lattice Boltzmann scheme [37], we adopt a different set of weight coefficients in Eq. (17) and Eq. (18) to correctly recover the special ratio of '1/3' between the terms $\nabla(\nabla \cdot \mathbf{q})$ and $\nabla^2 \mathbf{q}$ in the phonon hydrodynamic equation (7).

To sum up, we obtain a lattice Boltzmann equation (13) with the discrete equilibrium distribution Eq. (15) and the source term Eq. (16). Note the classical lattice Boltzmann equation can be directly derived from the BGK-Boltzmann equation using the Gauss-Hermite quadrature [38, 39] attributed to the Gaussian form ($\exp(-x^2)$) of the Maxwell-Boltzmann equilibrium distribution. As the Bose-Einstein equilibrium distribution Eq. (3) of phonons no longer has such a Gaussian form, it is difficult to construct the phonon lattice Boltzmann equation through a similar procedure.

Similar to the definition of momentum flux in classical lattice Boltzmann scheme with source term [40], the heat flux is defined as:

$$q_\alpha(\mathbf{x}, t) = \sum_i c_{i\alpha} e_i(\mathbf{x}, t) + \frac{1}{2} F_\alpha(\mathbf{x}, t) \Delta t, \qquad (19)$$

with the equivalent external force term: $F_\alpha = -q_\alpha / \tau_R$, which appears in the heat flux balance equation from the phonon resistive scattering. Similar to the treatment in classical lattice Boltzmann scheme with source term [37], Eq. (16) is treated as a first-order small quantity such that:

$$S_i(\mathbf{x}, t) = \varepsilon S_i^{(1)}(\mathbf{x}, t), \qquad (20)$$

$$F_\alpha(\mathbf{x}, t) = \varepsilon F_\alpha^{(1)}(\mathbf{x}, t), \qquad (21)$$

where the small parameter $\varepsilon$ will be explained in the following sub-section.

## 2.3. Chapman-Enskog expansion

To obtain the relation between the numerical parameters (lattice speed and non-dimensional relaxation time) and phonon properties (group speed and relaxation time), a Chapman-Enskog expansion [23, 24, 41] will be conducted to derive the phonon hydrodynamic equation (7) from the lattice Boltzmann equation (13).

For convenience of later derivation, the first four velocity moments of the discrete equilibrium energy density distribution are calculated as:

$$\sum_i e_i^{eq}(\mathbf{x}, t) = C_V T(\mathbf{x}, t), \qquad (22)$$

$$\sum_i c_{i\alpha} e_i^{\mathrm{eq}}(\mathbf{x}, t) = q_\alpha(\mathbf{x}, t), \tag{23}$$

$$\sum_i c_{i\alpha} c_{i\beta} e_i^{\mathrm{eq}}(\mathbf{x}, t) = \frac{5}{9} c^2 C_v T(\mathbf{x}, t) \delta_{\alpha\beta}, \tag{24}$$

$$\sum_i c_{i\alpha} c_{i\beta} c_{i\gamma} e_i^{\mathrm{eq}}(\mathbf{x}, t) = \frac{1}{3} c^2 \left[ q_\alpha(\mathbf{x}, t) \delta_{\beta\gamma} + q_\beta(\mathbf{x}, t) \delta_{\alpha\gamma} + q_\gamma(\mathbf{x}, t) \delta_{\alpha\beta} \right], \tag{25}$$

with $\delta_{\alpha\beta}$ the Kronecker delta function. Eq. (22) represents the energy conservation principle for both phonon resistive scattering and normal scattering, whereas Eq. (23) represents the quasi-momentum conservation principle for phonon normal scattering. These two conservation conditions have been important inspiration to construct the discrete equilibrium distribution Eq. (15) from the continuous one Eq. (5). With the help of Eq. (20) and Eq. (21), we derive the following relations from the first three velocity moments of the discrete source term:

$$\sum_i S_i^{(1)}(\mathbf{x}, t) = 0, \tag{26}$$

$$\sum_i c_{i\alpha} S_i^{(1)}(\mathbf{x}, t) = \left(1 - \frac{1}{2\tau}\right) F_\alpha^{(1)}(\mathbf{x}, t), \tag{27}$$

$$\sum_i c_{i\alpha} c_{i\beta} S_i^{(1)}(\mathbf{x}, t) = 0. \tag{28}$$

Following the standard procedures in the lattice Boltzmann method [23, 24, 41], we introduce the multiscale Chapman-Enskog expansion as:

$$e_i(\mathbf{x}, t) = e_i^{(0)}(\mathbf{x}, t) + \varepsilon e_i^{(1)}(\mathbf{x}, t) + \cdots, \tag{29}$$

$$\frac{\partial}{\partial t} = \varepsilon \frac{\partial}{\partial t_1} + \varepsilon^2 \frac{\partial}{\partial t_2}, \tag{30}$$

$$\frac{\partial}{\partial x_\alpha} = \varepsilon \frac{\partial}{\partial x_{1\alpha}}, \tag{31}$$

where $t_1$ and $t_2$ are respectively the time scale of advection and diffusion, whereas $\mathbf{x}_1$ denotes the spatial scale of both advection and diffusion. The zeroth-order distribution $e_i^{(0)}(\mathbf{x}, t)$ shall be the discrete equilibrium distribution Eq. (15). The small parameter $\varepsilon$ is explained as below. As is known in the kinetic theory of transport processes, through a non-dimensionalization of the Boltzmann equation, there will be a Knudsen number in the denominator of the right-hand collision term [42]. In the limit of small Knudsen number, one can derive the hydrodynamic equation of Boltzmann equation through the Chapman-Enskog expansion. Thus the small parameter $\varepsilon$ can be interpreted as the overall Knudsen number ($Kn_C$, defined as the ratio of overall phonon mean free path to characteristic length). The derived phonon hydrodynamic equation is inherently valid for small $Kn_C$, or equivalently, for characteristic time scale much larger than the overall phonon relaxation time.

As the first-order distribution has no contribution to the local energy density, one gets the following compatibility condition:

$$\sum_i e_i^{(1)}(\mathbf{x}, t) = 0.$$  (32)

From the calculation of heat flux Eq. (19), combined with Eq. (21) and Eq. (23), one gets another compatibility condition as:

$$\sum_i c_{i\alpha} e_i^{(1)}(\mathbf{x}, t) = -\frac{1}{2}\Delta t F_\alpha^{(1)}(\mathbf{x}, t).$$  (33)

Substituting Eqs. (29)-(31) into the phonon lattice Boltzmann equation (13) and collecting all the first-order terms ($O(\varepsilon^1)$), we obtain:

$$\frac{\partial e_i^{(0)}(\mathbf{x}, t)}{\partial t_1} + c_{i\alpha} \frac{\partial e_i^{(0)}(\mathbf{x}, t)}{\partial x_{1\alpha}} = -\frac{1}{\tau \Delta t} e_i^{(1)}(\mathbf{x}, t) + S_i^{(1)}(\mathbf{x}, t).$$  (34)

Taking the first two velocity moments of Eq. (34), we obtain:

$$\frac{\partial}{\partial t_1} \sum_i e_i^{(0)}(\mathbf{x}, t) + \frac{\partial}{\partial x_{1\alpha}} \sum_i c_{i\alpha} e_i^{(0)}(\mathbf{x}, t) = -\frac{1}{\tau \Delta t} \sum_i e_i^{(1)}(\mathbf{x}, t) + \sum_i S_i^{(1)}(\mathbf{x}, t),$$  (35)

$$\frac{\partial}{\partial t_1} \sum_i c_{i\alpha} e_i^{(0)}(\mathbf{x}, t) + \frac{\partial}{\partial x_{1\beta}} \sum_i c_{i\alpha} c_{i\beta} e_i^{(0)}(\mathbf{x}, t) = -\frac{1}{\tau \Delta t} \sum_i c_{i\alpha} e_i^{(1)}(\mathbf{x}, t) + \sum_i c_{i\alpha} S_i^{(1)}(\mathbf{x}, t).$$  (36)

With the help of Eqs. (22)-(24), Eq. (26) and Eq. (27), Eq. (32) and Eq. (33), Eq. (35) and Eq. (36) become:

$$C_V \frac{\partial T(\mathbf{x}, t)}{\partial t_1} + \frac{\partial q_\alpha(\mathbf{x}, t)}{\partial x_{1\alpha}} = 0,$$  (37)

$$\frac{\partial q_\alpha(\mathbf{x}, t)}{\partial t_1} + \frac{5}{9} c^2 C_V \frac{\partial T(\mathbf{x}, t)}{\partial x_{1\alpha}} = F_\alpha^{(1)}(\mathbf{x}, t).$$  (38)

Then we proceed to the next order. Substituting Eqs. (29)-(31) into the phonon lattice Boltzmann equation (13), expanding $e_i(\mathbf{x} + \mathbf{c}_i \Delta t, t + \Delta t)$ to second order and collecting all the second-order terms ($O(\varepsilon^2)$), we obtain:

$$\frac{\partial e_i^{(0)}(\mathbf{x}, t)}{\partial t_2} + \frac{\partial e_i^{(1)}(\mathbf{x}, t)}{\partial t_1} + c_{i\alpha} \frac{\partial e_i^{(1)}(\mathbf{x}, t)}{\partial x_{1\alpha}} + \frac{1}{2}\Delta t \left[ \frac{\partial^2 e_i^{(0)}(\mathbf{x}, t)}{\partial t_1 \partial t_1} + 2 c_{i\alpha} \frac{\partial^2 e_i^{(0)}(\mathbf{x}, t)}{\partial t_1 \partial x_{1\alpha}} + c_{i\alpha} c_{i\beta} \frac{\partial^2 e_i^{(0)}(\mathbf{x}, t)}{\partial x_{1\alpha} \partial x_{1\beta}} \right] = 0.$$

(39)

Taking the first two velocity moments of Eq. (39), we obtain:

$$\frac{\partial}{\partial t_2}\sum_i e_i^{(0)}(\mathbf{x},t)+\frac{\partial}{\partial t_1}\sum_i e_i^{(1)}(\mathbf{x},t)+\frac{\partial}{\partial x_{i\alpha}}\sum_i c_{i\alpha}e_i^{(1)}(\mathbf{x},t)$$
$$+\frac{1}{2}\Delta t\left[\frac{\partial^2}{\partial t_1\partial t_1}\sum_i e_i^{(0)}(\mathbf{x},t)+2\frac{\partial^2}{\partial t_1\partial x_{i\alpha}}\sum_i c_{i\alpha}e_i^{(0)}(\mathbf{x},t)+\frac{\partial^2}{\partial x_{1\alpha}\partial x_{1\beta}}\sum_i c_{i\alpha}c_{i\beta}e_i^{(0)}(\mathbf{x},t)\right]=0 \tag{40}$$

$$\frac{\partial}{\partial t_2}\sum_i c_{i\alpha}e_i^{(0)}(\mathbf{x},t)+\frac{\partial}{\partial t_1}\sum_i c_{i\alpha}e_i^{(1)}(\mathbf{x},t)+\frac{\partial}{\partial x_{i\beta}}\sum_i c_{i\alpha}c_{i\beta}e_i^{(1)}(\mathbf{x},t)$$
$$+\frac{1}{2}\Delta t\left[\frac{\partial^2}{\partial t_1\partial t_1}\sum_i c_{i\alpha}e_i^{(0)}(\mathbf{x},t)+2\frac{\partial^2}{\partial t_1\partial x_{1\beta}}\sum_i c_{i\alpha}c_{i\beta}e_i^{(0)}(\mathbf{x},t)+\frac{\partial^2}{\partial x_{1\beta}\partial x_{1\gamma}}\sum_i c_{i\alpha}c_{i\beta}c_{i\gamma}e_i^{(0)}(\mathbf{x},t)\right]=0 \tag{41}$$

With the help of Eqs. (22)-(25), Eq. (32) and Eq. (33), as well as the first-order equations (37) and (38), Eq. (40) and Eq. (41) become:

$$C_V\frac{\partial T(\mathbf{x},t)}{\partial t_2}=0 , \tag{42}$$

$$\frac{\partial q_\alpha(\mathbf{x},t)}{\partial t_2}+\frac{\partial}{\partial x_{1\beta}}\sum_i c_{i\alpha}c_{i\beta}e_i^{(1)}(\mathbf{x},t)+\frac{1}{6}\Delta tc^2\left[\frac{\partial^2 q_\alpha(\mathbf{x},t)}{\partial x_{1\beta}\partial x_{1\beta}}+\frac{1}{3}\frac{\partial}{\partial x_{1\alpha}}\left(\frac{\partial q_\beta(\mathbf{x},t)}{\partial x_{1\beta}}\right)\right]=0 . \tag{43}$$

The first-order approximation of discrete energy density distribution is related to the zeroth-order one through Eq. (34):

$$e_i^{(1)}(\mathbf{x},t)=\tau\Delta tS_i^{(1)}(\mathbf{x},t)-\tau\Delta t\frac{\partial e_i^{(0)}(\mathbf{x},t)}{\partial t_1}-\tau\Delta tc_{i\alpha}\frac{\partial e_i^{(0)}(\mathbf{x},t)}{\partial x_{1\alpha}} . \tag{44}$$

From Eq. (44), we obtain the second-order velocity moment of the first-order discrete energy density distribution as:

$$\sum_i c_{i\alpha}c_{i\beta}e_i^{(1)}(\mathbf{x},t)=\frac{2}{9}\tau\Delta tc^2\frac{\partial q_\gamma(\mathbf{x},t)}{\partial x_{1\gamma}}\delta_{\alpha\beta}-\frac{1}{3}\tau\Delta tc^2\left[\frac{\partial q_\alpha(\mathbf{x},t)}{\partial x_{1\beta}}+\frac{\partial q_\beta(\mathbf{x},t)}{\partial x_{1\alpha}}\right]. \tag{45}$$

Putting Eq. (45) into Eq. (43), we obtain the second-order equation as:

$$\frac{\partial q_\alpha(\mathbf{x},t)}{\partial t_2}=\left(1-\frac{1}{2\tau}\right)\frac{1}{3}\tau\Delta tc^2\left[\frac{\partial^2 q_\alpha(\mathbf{x},t)}{\partial x_{1\beta}\partial x_{1\beta}}+\frac{1}{3}\frac{\partial}{\partial x_{1\alpha}}\left(\frac{\partial q_\beta(\mathbf{x},t)}{\partial x_{1\beta}}\right)\right]. \tag{46}$$

Finally summing over the four equations in first order and second order, we obtain the energy balance equation (6) and the following heat transport equation:

$$\frac{\partial q_\alpha}{\partial t}+\frac{5}{9}c^2C_V\frac{\partial T}{\partial x_\alpha}=-\frac{q_\alpha}{\tau_R}+\left(1-\frac{1}{2\tau}\right)\frac{1}{3}\tau\Delta tc^2\left[\frac{\partial^2 q_\alpha}{\partial x_\beta\partial x_\beta}+\frac{1}{3}\frac{\partial}{\partial x_\alpha}\left(\frac{\partial q_\beta}{\partial x_\beta}\right)\right], \tag{47}$$

where the notation of spatial and temporal dependence has been omitted for elegance. Eq. (47) is exactly the phonon hydrodynamic equation (7). Through an identification of the corresponding coefficients, we obtain the lattice speed and non-dimensional relaxation time respectively:

$$c = \sqrt{\frac{3}{5}} v_g, \tag{48}$$

$$\tau = \frac{\tau_C}{\Delta t} + \frac{1}{2}. \tag{49}$$

The lattice speed is related to rather than identical to the phonon group speed. The "1/2" in Eq. (49) comes from the discretization error of lattice Boltzmann scheme [43]. Through the Chapman-Enskog expansion, we provide a rigorous relation between the numerical parameters and the phonon properties of materials. As a result, the evolution of the phonon lattice Boltzmann equation (13) will correspond to the numerical solution of the phonon hydrodynamic equation (7). This holds the similar logic to that the evolution of classical lattice Boltzmann equation corresponds to the numerical solution of Navier-Stokes hydrodynamic equation [23, 24]. Nevertheless, there are two essential differences in the phonon hydrodynamic equation from the classical hydrodynamic equation: (***i***) the lack of spatial inertia term like $\mathbf{q} \cdot \nabla \mathbf{q}$, the underlying mechanism of which requires further exploration and, (***ii***) the additional intrinsic source term due to the phonon resistive scattering. Thus the macroscopic hydrodynamic equation (7) or (47) is not a normal advection-diffusion equation in classical lattice Boltzmann scheme. Note the present phonon lattice Boltzmann scheme satisfies the Courant-Friedrichs-Lewy (CFL) stability condition as explained as follows. The phonon lattice Boltzmann equation (13) can be treated as a special finite difference approximation (explicit time and upwind spatial discretization) of the following partial differential equation:

$$\frac{\partial e_i(\mathbf{x}, t)}{\partial t} + \mathbf{c}_i \cdot \nabla e_i(\mathbf{x}, t) = \frac{1}{\tau} \left[ e_i^{\text{eq}}(\mathbf{x}, t) - e_i(\mathbf{x}, t) \right] + \Delta t S_i(\mathbf{x}, t). \tag{50}$$

For the present two-dimensional lattice Boltzmann scheme, the CFL condition is:

$$\frac{|c_{ix}| \Delta t}{\Delta x} \le 1, \frac{|c_{iy}| \Delta t}{\Delta y} \le 1. \tag{51}$$

As $\Delta x = \Delta y = c \Delta t$, with also the help of the value of discrete lattice velocities in Eq. (14), the CFL condition in Eq. (51) is satisfied. Eq. (48) relates the lattice speed $c$ to the phonon group velocity $v_g$ as a material parameter, and the velocity $\mathbf{c}_i$ in the partial differential equation (50) rather than the group velocity $v_g$ shall be adopted in the calculation of CFL number.

In comparison to the previous lattice Boltzmann scheme for the Callaway model [25], the present scheme has several crucial improvements: (*i*) the lattice speed is assumed from the outset as the phonon group speed [25] and will induce a fictitious phonon speed along the diagonal direction of a square lattice system [29, 44], whereas the lattice speed is related to the group speed as in Eq. (48) in the present scheme; (*ii*) the weight coefficients have a complex dependence on both the lattice speed and the derivative of Planck distribution [25], whereas constant numeric values are adopted in Eq. (17) and Eq. (18) in the present scheme. Because of the two points, the phonon hydrodynamic equations have not been correctly recovered [25], which can be further inferred from the following two aspects: (*i*) in the limit of dominant phonon normal scattering, the derived heat transport equation cannot be reduced exactly to the G-K equation [25]; (*ii*) there is an additional fourth-order spatial derivative term in the temperature differential equation [25]. In contrast, the phonon hydrodynamic equation (7) or (47) in the present scheme is reduced to the G-K equation (8) in the limit of dominant normal scattering. On the other hand, there is no longer the nonphysical fourth-order derivative term in the present temperature differential equation obtained by combining Eq. (6) and Eq. (7):

$$\frac{\partial^2 T}{\partial t^2} + \frac{1}{\tau_R}\frac{\partial T}{\partial t} = \frac{1}{3}v_g^2\nabla^2 T + \frac{4}{15}\tau_C v_g^2\frac{\partial}{\partial t}\left(\nabla^2 T\right). \tag{52}$$

In addition, although the relation between an effective phonon mean free path and the relaxation time of Umklapp scattering in the lattice Boltzmann equation has been discussed [28], the extraction of this physical mean free path is based on a phenomenological expression of size-dependent effective thermal conductivity and a one-dimensional heat transport process.

Finally the applicability of the phonon lattice Boltzmann scheme is explained. Although the hydrodynamic equation (7) together with the boundary conditions (10) and (11) is able to describe phonon transport within the slip regime (as shown in Appendix A), the present lattice Boltzmann scheme is merely available for phonon transport in the near-continuum regime due to insufficient number of discrete lattice velocities. A similar logic holds in the classical lattice Boltzmann scheme [24]: the Navier-Stokes equation with slip boundary conditions is able to describe gas flow within slip regime, while the lattice Boltzmann scheme in its original form (without any modifications) is merely available for gas flow in near-continuum regime.

## 2.4. Macroscopic variables calculation

Once resolving the discrete energy density distribution in the phonon lattice Boltzmann equation (13), we can calculate the macroscopic field variables by taking the moments. The temperature distribution is computed as:

$$e(\mathbf{x},t) = \sum_i e_i(\mathbf{x},t) = C_V T(\mathbf{x},t). \tag{53}$$

The calculation of heat flux distribution Eq. (19) is slightly transformed into:

$$q_\alpha(\mathbf{x},t) = \frac{2\tau_r}{2\tau_r + 1} \sum_i c_{i\alpha} e_i(\mathbf{x},t), \tag{54}$$

where the non-dimensional relaxation time for phonon resistive scattering is: $\tau_r = \tau_R/\Delta t$.

## 2.5. Boundary treatments

In this sub-section, the boundary treatment in the lattice Boltzmann scheme for phonon hydrodynamic equation is introduced. Several common boundary conditions are considered: (**i**) isothermal boundary condition, (**ii**) heat flux boundary condition, (**iii**) adiabatic boundary condition, and (**iv**) periodic temperature gradient boundary condition. In terms of the isothermal boundary condition, the discrete energy density distribution is given as the following equilibrium distribution:

$$e_i(\mathbf{x}_s,t) = \omega_i C_V T(\mathbf{x}_s,t) \quad (i = 1,2,\cdots 8). \tag{55}$$

For the heat flux boundary condition, we specify a heat flux $q_0(\mathbf{x}_s,t)$ incident into a certain surface:

$$q_0(\mathbf{x}_s,t) = -\frac{2\tau_r}{2\tau_r + 1} \sum_i (\mathbf{c}_i \cdot \mathbf{n}) e_i(\mathbf{x}_s,t), \tag{56}$$

where $\mathbf{n}$ denotes the unit external normal vector at the surface. Based on Eq. (56), we derive the unknown components of discrete energy density distribution at the boundary nodes after the advection step:

$$e_i\left(\mathbf{x}_s,t\right)=\frac{\omega_i}{\sum_{\mathbf{c}_i\cdot\mathbf{n}<0}\omega_i}\left[\sum_{\mathbf{c}_i\cdot\mathbf{n}>0}e_i\left(\mathbf{x}_s,t\right)+\frac{q_0\left(\mathbf{x}_s,t\right)}{c}\cdot\frac{2\tau_r+1}{2\tau_r}\right],\ \mathbf{c}_i\cdot\mathbf{n}<0\,,\tag{57}$$

with $\mathbf{c}_i\cdot\mathbf{n}>0$ and $\mathbf{c}_i\cdot\mathbf{n}<0$ representing the phonons incident on the surface and leaving from the surface. The adiabatic boundary condition is implemented through a diffuse reflection scheme from the kinetic theory boundary conditions for discrete Boltzmann equations [45-47]:

$$e_i\left(\mathbf{x}_s,t\right)=\frac{\omega_i}{\sum_{\mathbf{c}_i\cdot\mathbf{n}<0}\omega_i}\sum_{\mathbf{c}_i\cdot\mathbf{n}>0}e_i\left(\mathbf{x}_s,t\right),\ \mathbf{c}_i\cdot\mathbf{n}<0\,.\tag{58}$$

The adiabatic boundary treatment Eq. (58) can be also viewed as a special case of the heat flux boundary treatment Eq. (57) at vanishing incident heat flux. The periodic temperature gradient boundary condition is introduced to exert a constant temperature gradient along a simulation domain [29]:

$$e_i\left(\mathrm{L}\right)=\omega_i C_V T_\mathrm{h}+e_i\left(\mathrm{R}\right)-\omega_i C_V T_\mathrm{c}\quad\left(i=1,5,8\right),\tag{59}$$

$$e_i\left(\mathrm{R}\right)=\omega_i C_V T_\mathrm{c}+e_i\left(\mathrm{L}\right)-\omega_i C_V T_\mathrm{h}\quad\left(i=3,6,7\right),\tag{60}$$

where 'L' and 'R' represents the left-hand side and right-hand side of the simulation domain. The periodic temperature gradient boundary condition is very crucial in the simulation of phonon transport in a periodic system, for instance, the in-plane phonon transport in Section 3.2.

# 3. Results and Discussion

In this section, we demonstrate a robust numerical validation of the developed lattice Boltzmann scheme for phonon hydrodynamic equation in Section 2. In Section 3.1, we firstly consider the phonon transport in diffusive limit where the resistive scattering dominates over the normal scattering such that the present modeling result will recover the solution of classical Fourier's law. Then the steady-state in-plane and cross-plane phonon transport through thin film is modeled in Section 3.2 and Section 3.3 from the regime where the phonon normal scattering dominates over the resistive scattering to the regime where both categories of phonon scattering are comparable. Finally in Section 3.4, the transient phonon heat pulse propagation is considered, with the entire transition from second sound to diffusive transport accurately captured by the lattice Boltzmann scheme. The overall Knudsen number is always smaller than 0.01 in the considered cases although either resistive Knudsen number or normal Knudsen number ($Kn_R$ or $Kn_N$, defined as the ratio of the phonon mean free path of resistive scattering or normal scattering to the characteristic length) can be very large ($1/Kn_C=1/Kn_R+1/Kn_N$). In other words, we simulate the phonon transport within the near-continuum regime by the present lattice Boltzmann scheme. For the convenience of studying the influence of relative strength of two phonon scattering processes on the phonon transport behaviors, a pseudo-material is adopted throughout the present work. In spite of using an artificial material, the main physical features in different transport regimes are captured quite well. In the future we will take into account the realistic material, for which the transport regime is mainly determined by the ratio of average linewidth (inverse of phonon relaxation time) of normal scattering to that of resistive scattering [12, 13]. This ratio of average linewidth varies from about $10^3$ at 30K to about 4 at 200K for isotopically pure graphene [13], and can vary between $1\sim10^6$ from room temperature to around 1K for bismuth [48]. The ratio of average phonon mean free path of resistive scattering to that of normal scattering lies within the scope of $1\sim100$ between 10K~25K for NaF with different purities [49]. The specific heat capacity and phonon group speed is chosen as $C_V=1.66\times10^6$(J/m$^3\cdot$K) and $v_g=6400$(m/s) respectively. For the phonon relaxation time, $\tau_R=6.53\times10^{-12}$(s) is adopted for heat transport in the diffusive regime in Section 3.1, whereas $\tau_N=6.53\times10^{-12}$(s) is adopted for other cases in Sections 3.2-3.4 where the resistive scattering rate is adjustable at a fixed normal scattering rate.

## 3.1. Phonon transport in diffusive limit

Two classical cases of heat transport in diffusive limit [29] are taken into account:

the one-dimensional transient phonon transport in Figure 2(a) and a two-dimensional steady-state phonon transport in Figure 2(b).

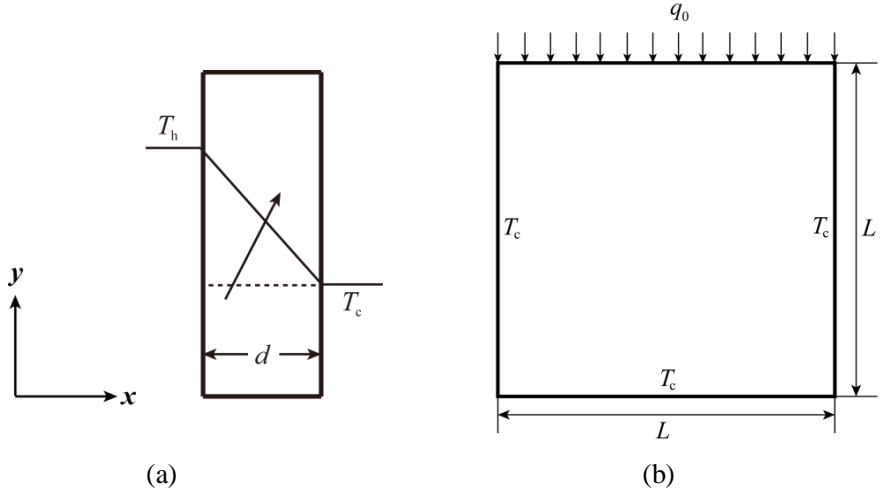

Figure 2. Schematic of physical models for heat transport in diffusive limit: (a) one-dimensional transient phonon transport; (b) two-dimensional steady-state phonon transport.

### 3.1.1 One-dimensional transient phonon transport

For the one-dimensional (1D) transient phonon transport across a thin film with a thickness $d$ in Figure 2(a), the thin film is initially at a uniform temperature $T_c$=299K. Suddenly, the left-hand side of the thin film keeps in contact with a hot source at a temperature $T_h$=301K whereas the right-hand side keeps in contact with a cold source at a temperature $T_c$. In the lattice Boltzmann modeling, the resistive and normal Knudsen numbers are set respectively as: $Kn_R$=1.29×10$^{-3}$ and $Kn_N$=1.29×10$^3$. The resistive scattering rate is six orders of magnitude stronger than the normal scattering rate such that the heat transport is dictated by the former. On the other hand, the resistive Knudsen number is sufficiently small to ensure the heat transport in diffusive regime. The D2Q8 lattice structure is used, with a grid of 201×3 adopted after an independence check. The left-hand and right-hand sides of the simulation domain are treated with the isothermal boundary conditions, whereas the upper and lower sides are treated with periodic boundary conditions to ensure a 1D transport.

The analytical solution of Fourier's heat equation for this 1D transient phonon transport has been obtained by the method of variable separation [2], with the temporal evolutions of temperature and heat flux distributions given respectively as

[29]:

$$\Theta = 1 - X - \frac{2}{\pi} \sum_{n=1}^{\infty} \frac{1}{n} \sin(n\pi X) \exp\left(-Kn_R^2 \frac{n^2\pi^2 t^*}{3}\right), \tag{61}$$

$$Q = 1 + 2\sum_{n=1}^{\infty} \cos(n\pi X) \exp\left(-Kn_R^2 \frac{n^2\pi^2 t^*}{3}\right), \tag{62}$$

where the non-dimensional spatial and temporal coordinates are defined separately as: $X \equiv x/d$, $t^* \equiv t/\tau_R$, and the non-dimensional temperature and heat flux are defined as: $\Theta \equiv \frac{T-T_c}{T_h-T_c}$, $Q \equiv \frac{q_x d}{\lambda(T_h-T_c)}$, with the bulk thermal conductivity being $\lambda = \frac{1}{3}C_V v_g^2 \tau_R$. The numerical results by the lattice Boltzmann scheme are compared to the analytical solutions of Fourier's heat equation in Figure 3. The very good agreement indicates that the present lattice Boltzmann model reduces to the classical Fourier's law in the current limit of dominant phonon resistive scattering.

To show the order of convergence of the present lattice Boltzmann scheme, we calculate the error norms of both the temperature and heat flux distributions at the same dimensionless time $t^*=1\times10^6$ using the number of grid points respectively: N$x$=201, 101, 51, 21. The error norms of the temperature and heat flux distributions are separately defined as [46]:

$$E_1(\Theta) = \frac{\sum_x |\Theta_{LBM} - \Theta_a|}{\sum_x |\Theta_a|}, \tag{63}$$

$$E_1(Q) = \frac{\sum_x |Q_{LBM} - Q_a|}{\sum_x |Q_a|}, \tag{64}$$

where the subscripts 'LBM' and 'a' denote the numerical results by the present lattice Boltzmann method and the analytical results respectively, with the summation taken over all the lattice. The error norms versus the inverse of grid number are shown in Figure 4, which demonstrates that the present numerical scheme has approximately a first order of convergence in spatial step (as well as time step).

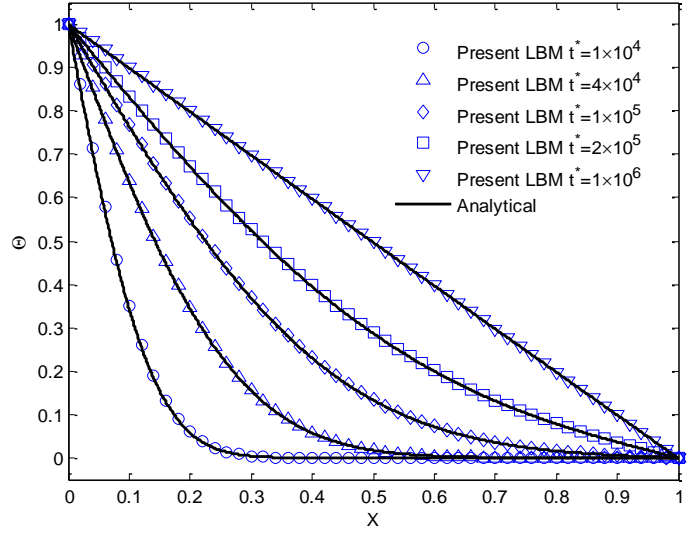

(a)

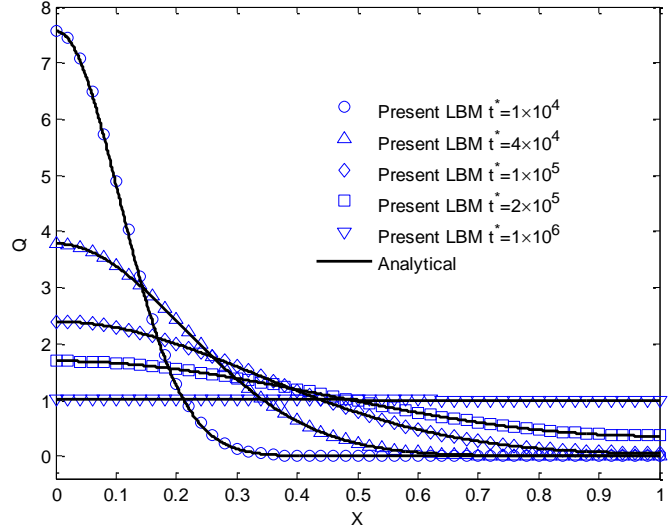

(b)

Figure 3. Temporal evolution of non-dimensional temperature and heat flux distributions in one-dimensional transient phonon transport in diffusive limit: (a) temperature, (b) heat flux, the symbols represent the numerical results by present LBM with $Kn_R=1.29\times10^{-3}$ and $Kn_N=1.29\times10^{3}$, whereas the solid lines represent the analytical solution Eq. (61) for temperature and Eq. (62) for heat flux of Fourier's heat equation. D2Q8 lattice structure is applied with a grid of 201×3 used after an independence check. The non-dimensional coordinates and variables are defined as: t*≡$t/\tau_R$, X≡$x/d$, Θ≡$(T-T_c)/(T_h-T_c)$, Q≡$q_x d/[\lambda(T_h-T_c)]$.

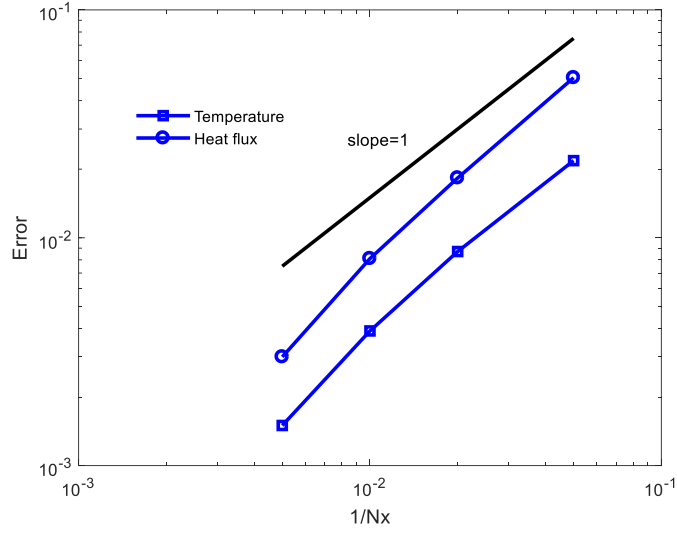

Figure 4. Error norm versus the inverse of the number of grid points in the lattice Boltzmann modeling of one-dimensional transient phonon transport, the line with squares and the line with circles represent respectively the results of temperature and heat flux distributions at the dimensionless time t*=1×10⁶.

### 3.1.2 Two-dimensional steady-state phonon transport

For the two-dimensional (2D) steady-state phonon transport in a square domain with a side length $L$ in Figure 2(b), a heat flux $q_0 = 1 \times 10^8$ (W/m²) is exerted on the top side whereas the other three sides keeps in contact with a cold source at a temperature $T_c$=299.5K. The resistive and normal Knudsen numbers are set as: $Kn_R$=2×10⁻³ and $Kn_N$=2×10³, which ensures the heat transport in diffusive regime. The D2Q8 lattice structure is used, with a grid 201×201 adopted after an independence check. The top side is treated with the heat flux boundary condition, whereas the other three sides are treated with isothermal boundary conditions. The analytical solution of temperature and heat flux distributions based on Fourier's heat equation for this 2D steady-state phonon transport has been obtained by the method of variable separation as [29]:

$$\Theta(X,Y) = \sum_{n=1}^{\infty} \frac{2Q_0}{(n\pi)^2} \frac{1-\cos n\pi}{\cosh n\pi} \sin(n\pi X) \sinh(n\pi Y),$$  (65)

$$Q_x(X,Y) = -\sum_{n=1}^{\infty} \frac{2Q_0}{n\pi} \frac{1-\cos n\pi}{\cosh n\pi} \cos(n\pi X) \sinh(n\pi Y),$$  (66)

$$Q_y(X, Y) = -\sum_{n=1}^{\infty} \frac{2Q_0}{n\pi} \frac{1 - \cos n\pi}{\cosh n\pi} \sin(n\pi X) \cosh(n\pi Y), \qquad (67)$$

where the non-dimensional spatial coordinates are defined as: $X \equiv x/L$, $Y \equiv y/L$, and the non-dimensional variables are defined as: $\Theta \equiv \dfrac{T - T_c}{T_c} \dfrac{1}{Kn_R}$, $Q \equiv \dfrac{qL}{\lambda T_c} \dfrac{1}{Kn_R}$. The numerical results of temperature distribution, $x$-direction heat flux distribution, $y$-direction heat flux distribution along three horizontal cross-sections of the square domain by the present lattice Boltzmann simulation are shown in Figure 5, which are well consistent with the Fourier's analytical solutions. The present case provides a good validation of the numerical treatment Eq. (57) for heat flux boundary condition.

The preceded two cases of heat transport in diffusive limit can be also described by the recent phonon lattice Boltzmann scheme under single mode relaxation time approximation [29], which considers only the resistive scattering. The present lattice Boltzmann scheme represents a further progress by taking into account the normal scattering based on the Callaway's dual relaxation model. In the following sections we will apply the present lattice Boltzmann scheme to explore the non-Fourier heat transport in hydrodynamic regime as well as in the transition from hydrodynamic regime to diffusive regime where the normal scattering plays an important role.

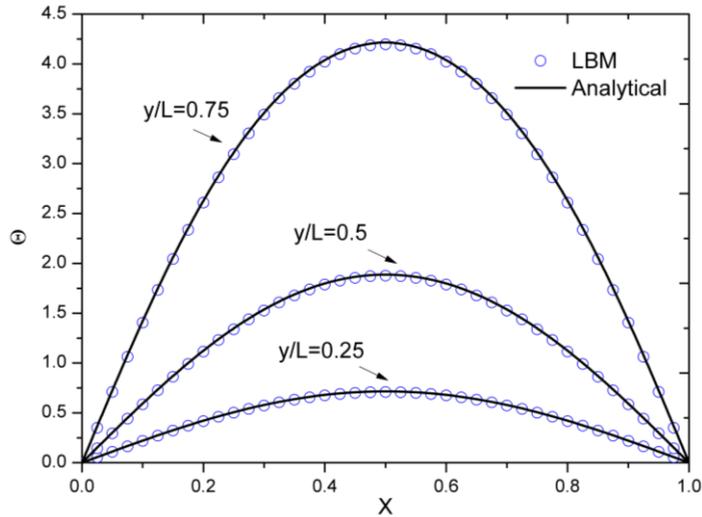

(a)

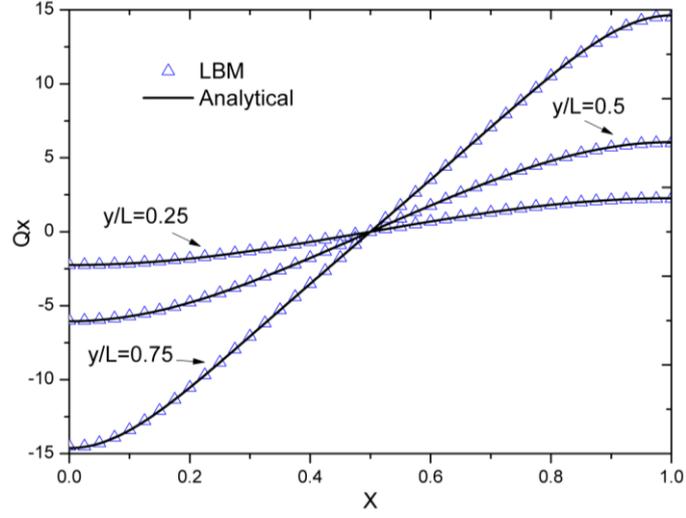

(b)

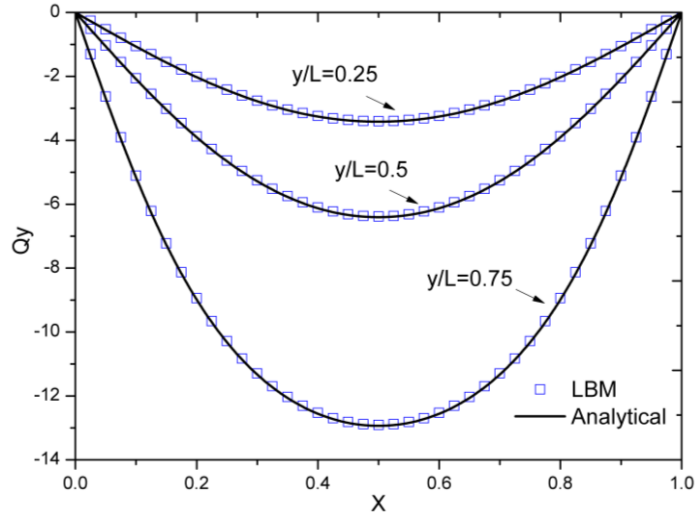

(c)

Figure 5. Non-dimensional temperature and heat flux distributions in two-dimensional steady-state phonon transport: (a) temperature, (b) $x$-direction heat flux, (c) $y$-direction heat flux along the cross-sections at $y/L$=0.25, 0.5, 0.75 respectively, the symbols represent the numerical results by present LBM with $Kn_R$=2×10⁻³ and $Kn_N$=2×10³, whereas the solid lines denote the analytical solution Eq. (65) for temperature, Eq. (66) for $x$-direction heat flux, Eq. (67) for $y$-direction heat flux of Fourier's heat equation. D2Q8 lattice structure is applied with a grid of 201×201 used after an independence check. The non-dimensional coordinates and variables are defined as: X≡$x/L$, Θ≡$(T-T_c)/T_c/Kn_R$, Q≡$qL/\lambda T_c/Kn_R$.

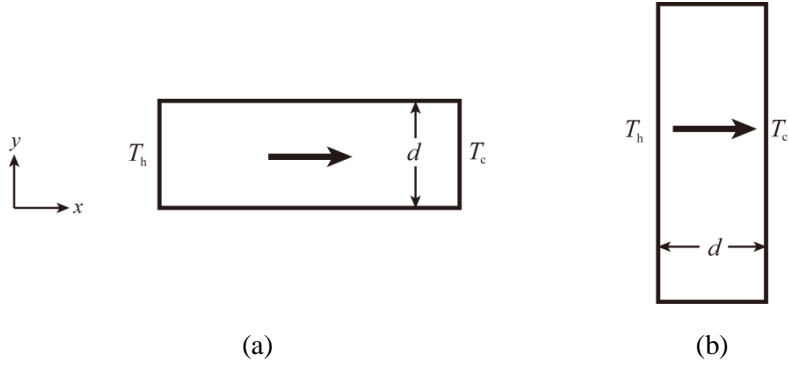

<center>(a)</center>

<center>(b)</center>

Figure 6. Schematic of (a) in-plane phonon transport and (b) cross-plane phonon transport

## 3.2. Steady-state in-plane phonon transport

The steady-state in-plane phonon transport through a thin film with a thickness $d$ is shown in Figure 6(a). The dimension of the in-plane direction is infinite such that there are no end effects. A constant temperature gradient is exerted along the thin film through the periodic temperature gradient boundary treatment. The normal Knudsen number is fixed at $Kn_N=0.01$, with the resistive Knudsen number varying from 0.01 to 1000. The D2Q8 lattice structure is used, with a grid 3×301 adopted after an independence check. The lateral adiabatic surfaces of the thin film are treated with the diffuse boundary condition. The analytical solution of heat flux distribution is obtained through solving the phonon hydrodynamic equation (7) with the heat flux boundary condition Eq.(11):

$$Q_x(Y)=1-\frac{\exp\left(\dfrac{Y-1}{Kn_{\text{eff}}}\right)+\exp\left(-\dfrac{Y}{Kn_{\text{eff}}}\right)}{1+\exp\left(-\dfrac{1}{Kn_{\text{eff}}}\right)+\dfrac{8}{15}\dfrac{Kn_C}{Kn_{\text{eff}}}\left[1-\exp\left(-\dfrac{1}{Kn_{\text{eff}}}\right)\right]}. \tag{68}$$

The effective Knudsen number in Eq. (68) is defined as $Kn_{\text{eff}}=\Lambda_{\text{eff}}/d$ based on an effective phonon mean free path: $\Lambda_{\text{eff}}=\sqrt{\tau_C\tau_R v_g^2/5}$, and the overall Knudsen number has been defined as $Kn_C=\Lambda_C/d$. The non-dimensional coordinate and heat flux is defined respectively as: $Y\equiv y/d$ and $Q_x(Y)\equiv q_x(y)\Big/\left(-\dfrac{1}{3}C_V v_g^2\tau_R\dfrac{dT}{dx}\right)$.

The numerical results by the present lattice Boltzmann scheme agree well with the analytical solutions as shown in Figure 7. When the normal scattering is absolutely dominant over the resistive scattering, the hydrodynamic phenomenon 'phonon Poiseuille flow' is obtained with parabolic heat flux profile, as seen for the case of $Kn_N$=0.01, $Kn_R$=100. With deceasing $Kn_R$, the phonon Poiseuille flow is gradually destroyed by the increasing momentum-destroying resistive scattering, which tends to introduce thermal resistance throughout the medium. Thus the heat flux profile is no longer parabolic as seen for the case of $Kn_N$=0.01, $Kn_R$=1. When the resistive scattering becomes comparable to the normal scattering, phonon transport becomes diffusive with a nearly uniform heat flux profile, as seen for the case of $Kn_N$=0.01, $Kn_R$=0.01. The present lattice Boltzmann model captures well the entire transition of heat transport behaviors from hydrodynamic regime to diffusive regime. This non-Fourier behavior induced by the normal scattering is very different from the ballistic-to-diffusive transition caused by boundary scattering in the usual in-plane phonon transport through thin film [4, 50], where the non-uniform heat flux profile in the ballistic limit is non-parabolic.

The effective in-plane thermal conductivity based on the phonon hydrodynamic model is computed through integrating the cross-sectional heat flux distribution Eq. (68):

$$\frac{\lambda_{\text{eff}}}{\lambda} = 1 - \frac{2Kn_{\text{eff}}\left[1 - \exp\left(-\frac{1}{Kn_{\text{eff}}}\right)\right]}{1 + \exp\left(-\frac{1}{Kn_{\text{eff}}}\right) + \frac{8}{15}\frac{Kn_{\text{C}}}{Kn_{\text{eff}}}\left[1 - \exp\left(-\frac{1}{Kn_{\text{eff}}}\right)\right]},$$ (69)

where the referenced thermal conductivity is $\lambda = \frac{1}{3}C_V v_g^2 \tau_R$, and the effective Knudsen number can be reformulated as $Kn_{\text{eff}} = \sqrt{Kn_{\text{C}} Kn_{\text{R}}/5}$. Since the heat flux profiles have been accurately predicted, the effective in-plane thermal conductivity by the present lattice Boltzmann scheme shows also a good agreement with the analytical solution Eq. (69) in Figure 8. We have also included the classical Fuchs-Sondheimer (F-S) model for the usual in-plane phonon transport through thin film [4] for a comparison. The introduction of phonon normal scattering will reduce the thermal resistance and increase the effective thermal conductivity a lot. This is attributed to the momentum conserving nature of normal scattering in contrast to the resistive scattering which destroys phonon momentum. As the classical F-S model is derived from the phonon Boltzmann equation under single mode relaxation time approximation, the counterpart model for the phonon Boltzmann equation under dual relaxation time approximation remains to be developed. The solution of Fourier's law in Figure 8 corresponds to the diffusive limit with absolutely dominant resistive scattering over normal scattering.

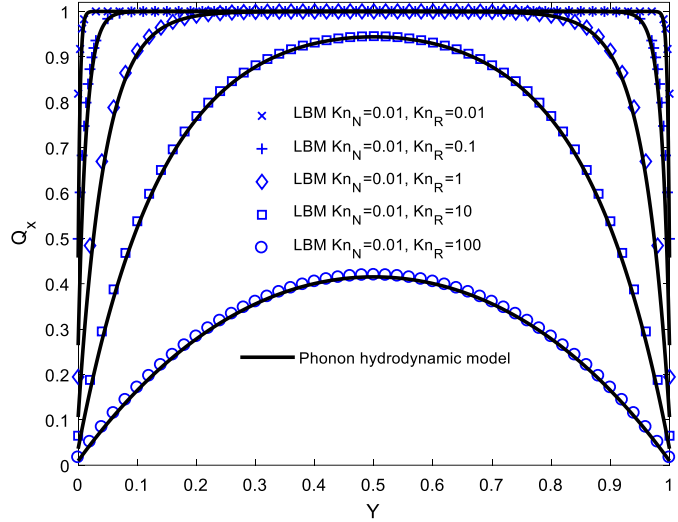

Figure 7. Non-dimensional cross-sectional heat flux distribution in in-plane phonon transport: the symbols represent the numerical results by the present LBM at $Kn_N=0.01$ and various $Kn_R$ from 0.01 to 100, whereas the solid lines represent the analytical solution Eq. (68) of phonon hydrodynamic model. The D2Q8 lattice structure is applied with a grid of 3×301 used after an independence check. The non-dimensional variables are respectively defined as: $Y\equiv y/d$ and $Qx\equiv-3q_x/(C_Vv_g\Lambda_R dT/dx)$.

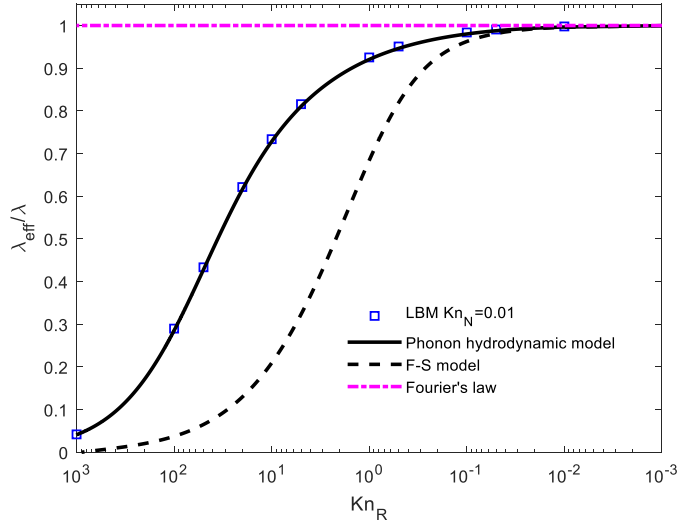

Figure 8. Non-dimensional effective in-plane thermal conductivity: the symbols represent the numerical results by the present LBM at $Kn_N=0.01$, the solid line, dashed line and dashed-dotted line represents the analytical solution Eq. (69) of phonon hydrodynamic model, the classical Fuchs-Sondheimer (F-S) model, and the Fourier's law respectively.

### 3.3. Steady-state cross-plane phonon transport

The steady-state cross-plane phonon transport through a thin film with a thickness $d$ is shown in Figure 6(b). The left-hand and right-hand sides of the thin film keep in contact with a hot source at a temperature $T_h$=301K and a cold source at a temperature $T_c$=299K respectively. The normal Knudsen number is fixed at $Kn_N$=0.01, with the resistive Knudsen number varying from 0.01 to 100. The D2Q8 lattice structure is used, with a grid 301×3 adopted after independence verification. The left-hand and right-hand sides of the simulation domain are treated with isothermal boundary conditions whereas the upper and lower sides are treated with periodic boundary condition to ensure one-dimensional transport. The analytical solutions of temperature distribution and heat flux are obtained through solving the phonon hydrodynamic equations (6) and (7) with the temperature jump boundary condition Eq. (10):

$$\Theta(X) = -\frac{X}{1+\frac{4}{3}Kn_R} + \frac{1+\frac{2}{3}Kn_R}{1+\frac{4}{3}Kn_R}, \tag{70}$$

$$Q = \frac{1}{3}\frac{Kn_R}{1+\frac{4}{3}Kn_R}. \tag{71}$$

The non-dimensional coordinate is defined as $X \equiv x/d$, whereas the non-dimensional temperature and heat flux are defined as: $\Theta \equiv \dfrac{T-T_c}{T_h-T_c}$ and $Q \equiv \dfrac{q_x}{C_v v_g (T_h - T_c)}$.

The numerical results by the present lattice Boltzmann scheme are in overall good agreement with the analytical solutions as shown in Figure 9. When the normal scattering dominates absolutely over the resistive scattering (as in the case $Kn_N$=0.01, $Kn_R$=100), the temperature distribution is nearly uniform inside the medium with a large jump at the boundary even in the near-continuum regime. This behavior is very different from the continuous linear temperature distribution in the usual cross-plane phonon transport through thin film in the diffusive limit [29, 51]. The underlying mechanism is attributed to the momentum conserving nature of normal scattering, which preserves a heat flow drift without thermal resistance [3]. Thus there is nearly no temperature drop across the medium. On the other hand, as the phonon distribution inside the medium is very close to the displaced Planck distribution, there exists a strong non-equilibrium between the phonons (at Planck distribution) emitted from the isothermal boundary and the phonons incident on the boundary from the medium. These two points explain the large boundary temperature jump, as well as the minor deviation of the numerical results from the analytical solution due to the insufficient discrete directions in the present lattice Boltzmann scheme to accurately capture the strong non-equilibrium effect near the boundary. The minor error at larger $Kn_R$ comes from the treatment of the strong non-equilibrium boundary, instead of the inherent accuracy of the lattice Boltzmann scheme. This can be further confirmed from the results in Section 3.4, where the transient cross-plane heat transport at small $Kn_N$ but large $Kn_R$ under heat flux boundary can be accurately predicted by the present scheme. It still requires further improvement of the treatment of isothermal boundary in the presence of dominant normal scattering over resistive scattering. With increasing resistive scattering rate (as in the case $Kn_N=0.01$, $Kn_R=1$), the collective effect of normal scattering weakens and the boundary temperature jump decreases. When the resistive scattering becomes comparable to the normal scattering (as in the case $Kn_N=0.01$, $Kn_R=0.01$), the heat transport becomes diffusive with a continuous linear temperature profile obtained. The solution of Fourier's law in Figure 9 corresponds to the diffusive limit with absolutely dominant resistive scattering over normal scattering.

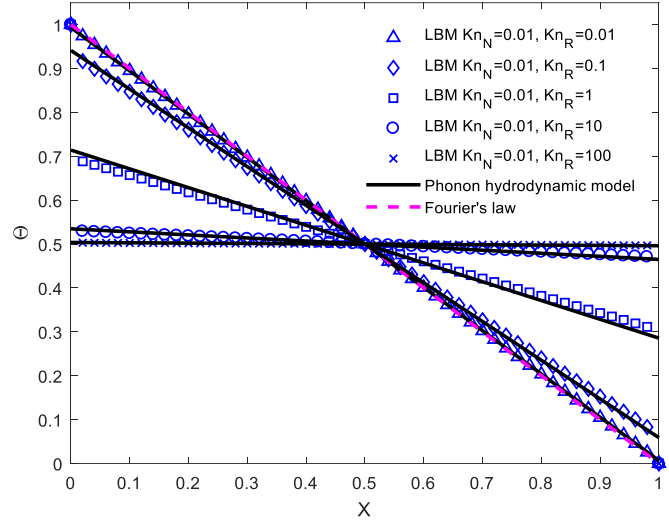

(a)

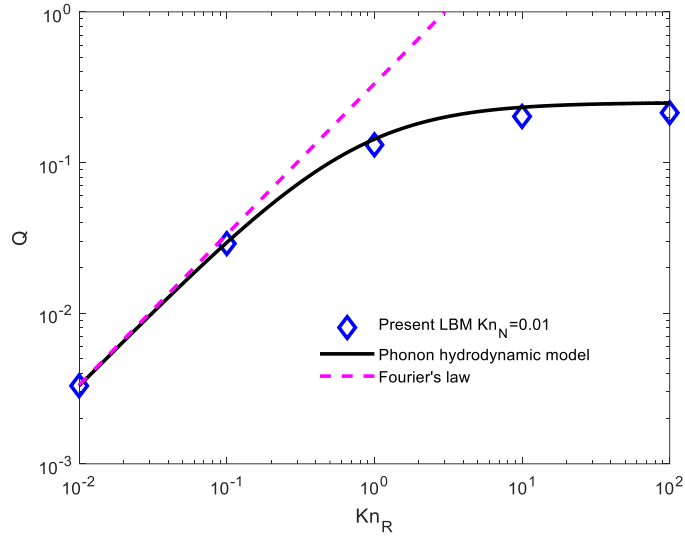

(b)

Figure 9. Non-dimensional temperature distribution and heat flux in cross-plane phonon transport: (a) temperature, (b) heat flux, the symbols represent the numerical results by the present LBM at $Kn_N$=0.01 and various $Kn_R$ from 0.01 to 100, whereas the solid lines and dashed line represent the analytical solution of phonon hydrodynamic model and Fourier's law respectively. D2Q8 lattice structure is applied with a grid of 301×3 used after an independence check. The non-dimensional variables are defined respectively as: X≡$x/d$, Θ≡$(T- T_c)/(T_h-T_c)$, Q≡$q_x/[C_V v_g(T_h-T_c)]$.

### 3.4. Transient phonon heat pulse propagation

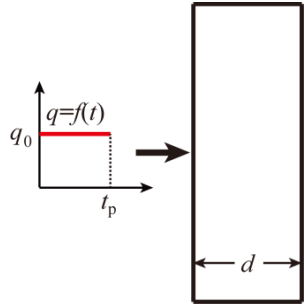

Figure 10. Schematic of transient heating of a plate by a short heat pulse with a magnitude $q_0$ and duration $t_p$

In this sub-section, we apply the present phonon lattice Boltzmann scheme to simulate the transient heat pulse propagation through a plate with a thickness $d$ as shown in Figure 10. The plate is initially at a temperature $T_0$. Suddenly, a heat pulse with a magnitude $q_0$ and a short duration $t_p$ is incident on the left-hand side of the plate. After the heat pulse is removed, the left-hand side of the plate keeps adiabatic. In comparison, the right-hand side of the plate keeps always adiabatic. This physical process is the paradigm for experimental detection of second sound in dielectric solid where phonon normal scattering is the dominant microscopic mechanism of heat transport [52, 53]. A direct numerical simulation of the second sound propagation has been considered mainly at the macroscopic level based on the nine-moment phonon hydrodynamic model [17], an empirical hybrid phonon gas hydrodynamic model [54], the dual-phase-lag model [6] or a nonequilibrium thermodynamic model with internal variables [55]. There is seldom report of a simulation of second sound based on a direct numerical solution of the phonon Boltzmann equation, *i.e.* at the mesoscopic level. The present mesoscopic scheme holds at least the following advantages over the macroscopic modeling: (***i***) a nature statistical physical foundation; (***ii***) a simpler solution of a discrete equation with linear advection term and algebraic collision term comparing to the numerical solution of partial differential equations. Comparing to the DOM scheme for the phonon Boltzmann equation under the Callaway model [13], the lattice Boltzmann scheme is more efficient in modeling transient heat transport attributed to avoiding internal iteration within each time step, and also facilitate modeling heat transport in complicated structures (such as porous structures) attributed to an easier treatment of complex boundary.

In this study, the normal Knudsen number is fixed at $Kn_N=1.29\times10^{-3}$. Three cases are considered with the resistive Knudsen number respectively as: $Kn_R=1.29\times10^3$, 1.29 and $1.29\times10^{-3}$. The initial temperature is set as $T_0=300$K. The magnitude and duration of the heat pulse is set as: $q_0 = 1\times10^8$ (W/m$^2$) and $t_p=100\tau_N$. The temperature range of second sound detection in the realistic three-dimensional dielectric materials is usually very low (around 10K) [53]. Although the temperature range in the present simulation is different from the value in real situation as a pseudo-material is adopted, the main features of heat pulse propagation in different transport regimes are the same. The D2Q8 lattice structure is used, with a grid 401×3 adopted after an independence check. The left-hand side of the simulation domain is treated with heat flux boundary condition during the pulse heating, and treated with diffuse boundary condition once the heat pulse is removed. The right-hand side of the simulation domain is treated with diffuse boundary condition. The upper and lower sides of the simulation domain are treated with periodic boundary condition to ensure one-dimensional transport.

An analytical solution of the phonon hydrodynamic model for this transient heat pulse propagation can be obtained as a benchmark. The energy balance equation and heat transport equation are reduced to:

$$C_V \frac{\partial T}{\partial t} + \frac{\partial q_x}{\partial x} = 0 \,, \tag{72}$$

$$\tau_R \frac{\partial q_x}{\partial t} + q_x = -\frac{1}{3} C_V v_g^2 \tau_R \frac{\partial T}{\partial x} + \frac{4}{15} \tau_C \tau_R v_g^2 \frac{\partial^2 q_x}{\partial x^2} \,. \tag{73}$$

The initial and boundary conditions are formulated respectively as:

$$t = 0, \ T(0 \le x \le d) = T_0, \ q_x(0 \le x \le d) = 0 \,, \tag{74}$$

$$\begin{cases} x = 0, \ q_x(t) = \Big[ H(t) - H\big(t - t_p\big) \Big] q_0 \\ x = d, \ q_x(t) = 0 \end{cases} \,, \tag{75}$$

where $H(t)$ is the Heaviside unit step function. For generality, the non-dimensional coordinates and variables are introduced as:

$$t^* \equiv \frac{t}{\tau_N}, \ X \equiv \frac{x}{d}, \ \Theta \equiv \frac{(T-T_0)C_V v_g}{q_0}, \ Q \equiv \frac{q_x}{q_0} \ . \tag{76}$$

In this way, we obtain the following non-dimensional governing equations:

$$\frac{\partial \Theta}{\partial t^*} + Kn_N \frac{\partial Q}{\partial X} = 0 \ , \tag{77}$$

$$\frac{Kn_R}{Kn_N} \frac{\partial Q}{\partial t^*} + Q = -\frac{1}{3} Kn_R \frac{\partial \Theta}{\partial X} + \frac{4}{15} Kn_C Kn_R \frac{\partial^2 Q}{\partial X^2} \ . \tag{78}$$

The non-dimensional initial and boundary conditions become:

$$t^* = 0, \ \Theta(0 \le X \le 1) = 0, \ Q(0 \le X \le 1) = 0 \ , \tag{79}$$

$$\begin{cases} X = 0, \ Q\left(t^*\right) = H\left(t^*\right) - H\left(t^* - 100\right) \\ X = 1, \ Q\left(t^*\right) = 0 \end{cases} . \tag{80}$$

Then the analytical solutions of the temporal evolution of temperature and heat flux distributions are obtained by the Laplace transform method (The mathematical details are provided in Appendix B).

A comparison of the temporal evolution of temperature and heat flux distributions between the lattice Boltzmann simulation and the phonon hydrodynamic modeling is shown in Figure 11 and Figure 12. A very good agreement is achieved for all the three cases. For the case of $Kn_N$=1.29×10$^{-3}$ and $Kn_R$=1.29×10$^3$ where the normal scattering absolutely dominates over the resistive scattering, the heat pulse propagates as a traveling wave known as the "second sound" as shown in Figure 11(a) and Figure 12(a). The heat pulse has merely a very small amount of dissipation due to the tiny portion of phonon resistive scattering. With increasing resistive scattering rate in the case of $Kn_N$=1.29×10$^{-3}$ and $Kn_R$=1.29, there exists significant dissipation of the second sound with nearly 50% reduction of the magnitude and about twofold broadening of the spatial width of heat pulse, as shown in Figure 11(b) and Figure 12(b). Multiple reflections take place when the second sound reaches the plate boundaries, which are captured well by the present numerical scheme. As the first arrival of the second sound is concerned in realistic experimental detection, we give merely the temporal evolutions till the second sound reaches the right-hand side of the plate for the first time. When the resistive scattering becomes comparable to the normal scattering in the case of $Kn_N$=1.29×10⁻³ and $Kn_R$=1.29×10⁻³, the second sound disappears and the heat pulse propagates in a diffusive way as shown in Figure 11(c) and Figure 12(c). The present lattice Boltzmann scheme describes well the transient heat transport behaviors from second sound propagation to diffusive transport. This transition is further illustrated in Figure 13, where the non-dimensional temperature responses at the right-hand side of the plate are given. When heat pulse propagates as a second sound, the right-hand side will receive a wave signal. The speed of the signal is computed through dividing the plate thickness by the arrival time, which is exactly consistent with the theoretical value: $v_g/\sqrt{3}$ [8, 17]. When heat pulse propagates in a diffusive way, the non-dimensional temperature of the right-hand side continuously increases and approaches the terminal rise, which is consistent with the exact value 0.129 calculated through dividing the total energy input of heat pulse by the system heat capacity.

Finally we give a brief discussion about the value of non-dimensional relaxation time $\tau$ in Eq. (49). In classical BGK lattice Boltzmann scheme, the non-dimensional relaxation time is related to the fluid viscosity $v$ as : $v = c_s^2 (\tau - 0.5)\Delta t$, with $c_s$ the sound speed and related to the lattice speed [23, 24]. The non-dimensional relaxation time $\tau$ has to satisfy $\tau > 0.5$ to ensure positive viscosity. On the other hand, $\tau$ cannot be too large for modeling of fluid flow in the near-continuum regime. As shown in Ref. [56], the BGK lattice Boltzmann scheme provides an accurate prediction when $0.2 < 1/\tau < 2$, or equivalently $0.5 < \tau < 5$. For liquid flow, a slightly smaller range of $0.5 < \tau < 2$ has been suggested [57]. For gas flow similar to the phonon transport case in this work, the non-dimensional relaxation time is directly related to the Knudsen number as [24]: $\tau = 0.5 + \sqrt{6/\pi} N_x Kn$, with $N_x$ the lattice number. For a typical Knudsen number $Kn$=0.01 in the near-continuum regime and a typical grid number $N_x$=100, one has $\tau$=1.8820. In the present phonon lattice Boltzmann scheme, we obtain a similar relation between the non-dimensional relaxation time and overall Knudsen number based on Eq. (48) and Eq. (49) as: $\tau = 0.5 + \sqrt{3/5} N_x Kn_C$. The maximum value of $\tau$ is evaluated to 0.6998, 0.8098, 2.8238 and 0.8997 for the 1D transient, 2D steady-state, steady-state in-plane or cross-plane, and transient heat pulse transport respectively. For the steady-state in-plane or cross-plane case, if we consider a smaller overall Knudsen number (closer to the continuum limit), we will obtain an even smaller $\tau$. In summary, the value of non-dimensional relaxation time in this work lies within a reasonable range. The good agreement between the numerical results and the analytical solution also validates the Chapman-Enskog analysis in Section 2.

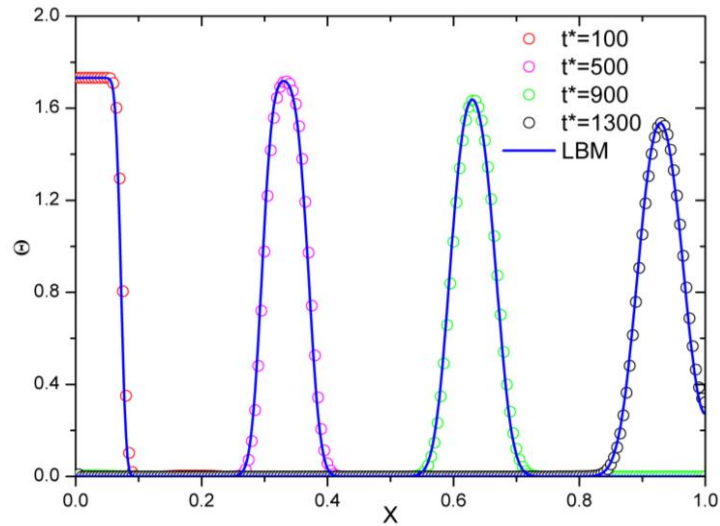

(a)

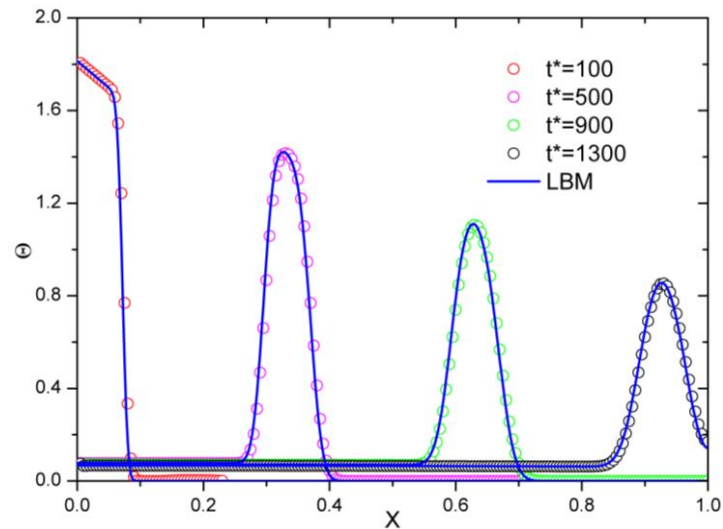

(b)

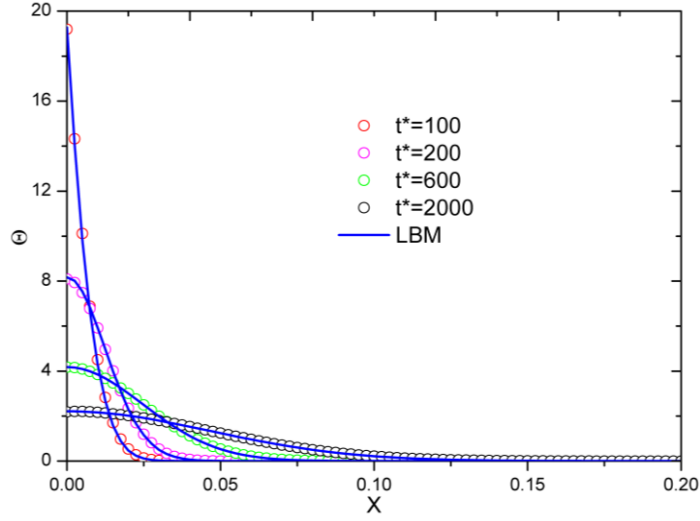

(c)

Figure 11. Temporal evolution of non-dimensional temperature distributions in transient phonon heat pulse propagation: (a) $Kn_N=1.29\times10^{-3}$ and $Kn_R=1.29\times10^3$; (b) $Kn_N=1.29\times10^{-3}$ and $Kn_R=1.29$; (c) $Kn_N=1.29\times10^{-3}$ and $Kn_R=1.29\times10^{-3}$. The symbols represent the analytical solution of the phonon hydrodynamic model by Laplace transform method, whereas the solid lines represent the numerical results by present LBM. D2Q8 lattice structure is applied with a grid of 401×3 used after an independence check. The non-dimensional variables are defined respectively as: $t^*\equiv t/\tau_N$, $X\equiv x/d$, $\Theta\equiv(T-T_0)C_Vv_g/q_0$. The magnitude and duration of heat pulse is $q_0=1\times10^8\text{W/m}^2$ and $t_p=100\tau_N$.

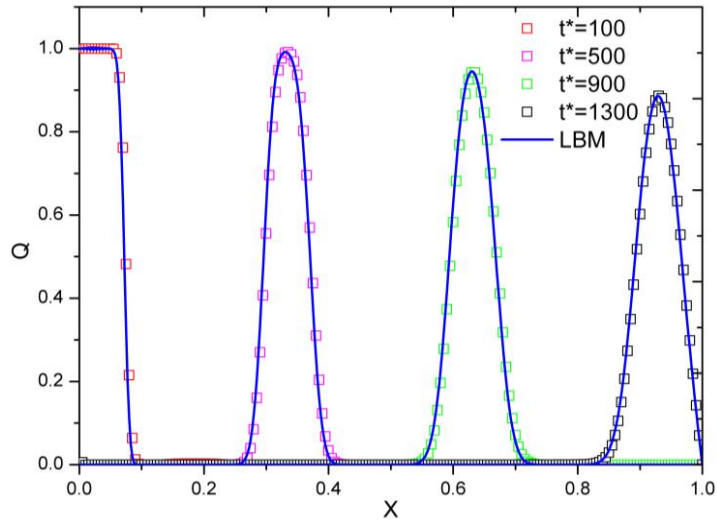

(a)

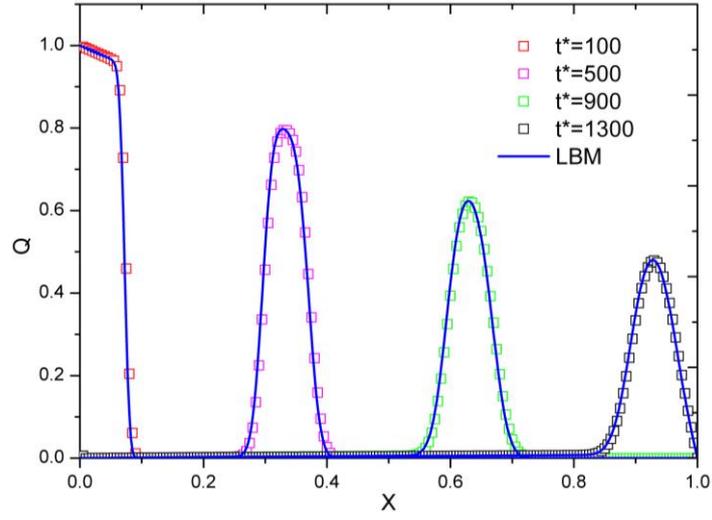

(b)

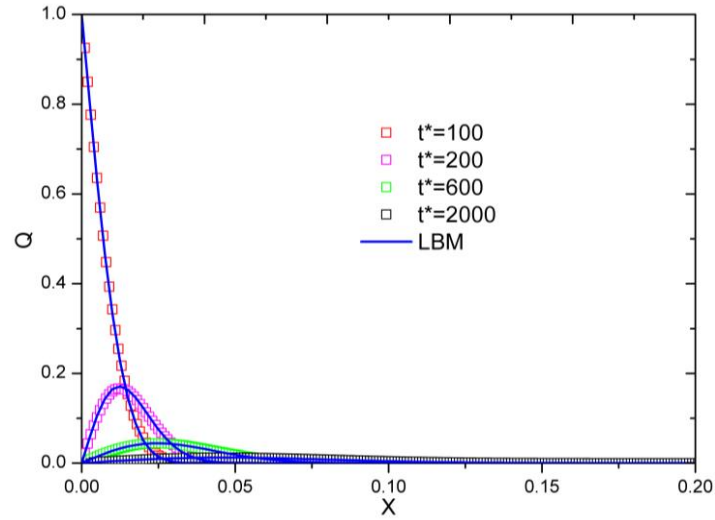

(c)

Figure 12. Temporal evolution of non-dimensional heat flux distributions in transient phonon heat pulse propagation: (a) $Kn_N=1.29\times10^{-3}$ and $Kn_R=1.29\times10^{3}$; (b) $Kn_N=1.29\times10^{-3}$ and $Kn_R=1.29$; (c) $Kn_N=1.29\times10^{-3}$ and $Kn_R=1.29\times10^{-3}$. The symbols represent the analytical solution of the phonon hydrodynamic model by Laplace transform method, whereas the solid lines represent the numerical results by present LBM. D2Q8 lattice structure is applied with a grid of 401×3 used after an independence check. The non-dimensional variables are defined respectively as: t*≡$t/\tau_N$, X≡$x/d$, Q≡$q_x/q_0$. The magnitude and duration of heat pulse is $q_0=1\times10^{8}$W/m² and $t_P=100\tau_N$.

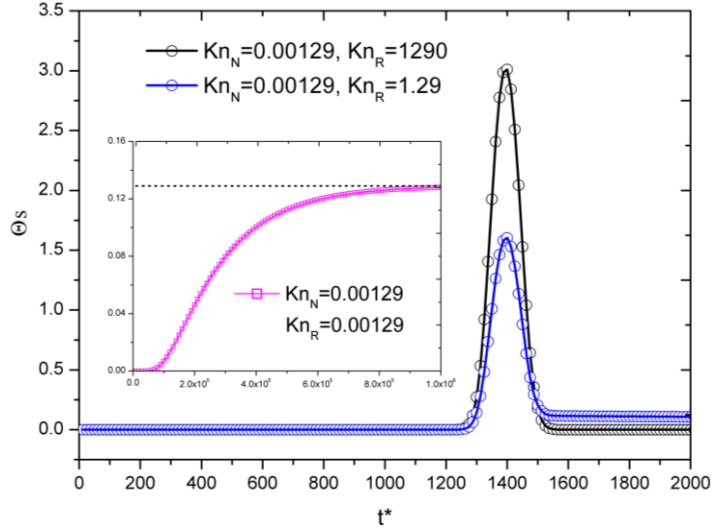

Figure 13. The non-dimensional temperature response of right-hand surface in transient phonon heat pulse propagation: the lines with circles represent the numerical results by present LBM for the case of $Kn_N$=1.29×10⁻³, $Kn_R$=1.29×10³ and $Kn_N$=1.29×10⁻³, $Kn_R$=1.29, whereas the line with squares in the inset figure denotes the numerical results by present LBM for the case of $Kn_N$=1.29×10⁻³, $Kn_R$=1.29×10⁻³, where the dashed line represent the final non-dimensional temperature rise.

## 4. Conclusions

A lattice Boltzmann scheme is established for the hydrodynamic equations of the phonon Boltzmann equation under Callaway's dual relaxation model in the continuum limit. The numerical parameters are related to the phonon properties rigorously through a Chapman-Enskog expansion to the lattice Boltzmann equation by treating the resistive scattering term as an equivalent source term. The recovered phonon hydrodynamic equation is reduced to the Guyer-Krumhansl heat transport equation and Fourier's law in the opposite limits of dominate normal scattering and dominant resistive scattering respectively. A robust validation of the numerical scheme is made for phonon heat transport from the hydrodynamic regime to diffusive regime. Two well-known phonon hydrodynamic phenomena are accurately captured including the phonon Poiseuille flow and second sound propagation. The non-Fourier heat transport behaviors in the hydrodynamic-to-diffusive transition are very different from that in the ballistic-to-diffusive transition of nanoscale heat transport. The lattice Boltzmann scheme provides a credible approach to the manipulation and optimization of hydrodynamic phonon transport in the future. The present work will promote not only the mesoscopic numerical modeling but also a deeper understanding of non-Fourier heat conduction in the presence of phonon normal scattering.

## Acknowledgements

This work is financially supported by the NSF grant of China (No. 91634107, 51621062).

# Appendix A. Derivation and validation of the phonon hydrodynamic equations

The macroscopic moment balance equations are derived through taking the moments of the phonon Boltzmann equation (1) [17]:

$$\frac{\partial e}{\partial t} + \frac{\partial q_\alpha}{\partial x_\alpha} = 0, \tag{A1}$$

$$\frac{\partial q_\alpha}{\partial t} + \frac{\partial Q_{\alpha\beta}}{\partial x_\beta} = -\frac{q_\alpha}{\tau_R}, \tag{A2}$$

$$\frac{\partial Q_{\alpha\beta}}{\partial t} + \frac{\partial M_{\alpha\beta\gamma}}{\partial x_\gamma} = \frac{1}{\tau_C}\left(\frac{1}{3}v_g^2 e\delta_{\alpha\beta} - Q_{\alpha\beta}\right). \tag{A3}$$

The energy density and heat flux are related to the phonon distribution function as: $e = \int \hbar\omega f d\mathbf{k}$ and $q_\alpha = \int v_{g\alpha}\hbar\omega f d\mathbf{k}$. The flux of heat flux and its higher-order flux are defined as: $Q_{\alpha\beta} = \int v_{g\alpha}v_{g\beta}\hbar\omega f d\mathbf{k}$ and $M_{\alpha\beta\gamma} = \int v_{g\alpha}v_{g\beta}v_{g\gamma}\hbar\omega f d\mathbf{k}$ respectively. To get a closed description of phonon heat transport, the higher-order fluxes have to be expressed as a function of the basic field variables (temperature and heat flux), which constitutes the closure problem in kinetic theory.

In a recent work [30], the regularized moment method in rarefied gas flow [58] has been adapted to derive a phonon hydrodynamic equation for nanoscale heat transport at ordinary temperatures from the phonon Boltzmann equation under single mode relaxation time approximation. In the present study, we extend this closure method to phonon transport based on Callaway's dual relaxation model. The main idea of the regularized moment method is a perturbation expansion around a referenced nonequilibrium distribution rather than the usual equilibrium one [58]. The referenced four moment non-equilibrium phonon distribution is obtained by the maximum entropy principle [17, 30]:

$$f_4 = f_R^{eq} + \frac{3}{C_V v_g^2}\frac{\partial f_R^{eq}}{\partial T}q_\alpha v_{g\alpha}. \tag{A4}$$

A perturbation expansion of $Q_{\alpha\beta} = Q_{\alpha\beta}^{(0)} + \varepsilon Q_{\alpha\beta}^{(1)} + \cdots$ is substituted in to Eq. (A3), which gives rise to $Q_{\alpha\beta}^{(0)} = \dfrac{1}{3} v_g^2 e \delta_{\alpha\beta}$ corresponding to $f_4$ and the first-order approximation:

$$Q_{\alpha\beta}^{(1)} = -\tau_C \left[ \frac{\partial}{\partial t} \left( Q_{\alpha\beta}^{(0)} \big|_{f_4} \right) + \frac{\partial}{\partial x_\gamma} \left( M_{\alpha\beta\gamma}^{(0)} \big|_{f_4} \right) \right]. \tag{A5}$$

The small parameter $\varepsilon$ here is interpreted as the overall Knudsen number, the same as

that in the main text. Putting Eq. (A4) into Eq. (A5) and combining the zeroth-order

approximation, we obtain the flux of heat flux:

$$Q_{\alpha\beta} = \frac{1}{3} v_g^2 e \delta_{\alpha\beta} + \frac{2}{15} \tau_C v_g^2 \frac{\partial q_\gamma}{\partial x_\gamma} \delta_{\alpha\beta} - \frac{1}{5} \tau_C v_g^2 \left( \frac{\partial q_\alpha}{\partial x_\beta} + \frac{\partial q_\beta}{\partial x_\alpha} \right). \tag{A6}$$

Therefore, we acquire the phonon hydrodynamic equation after substituting Eq. (A6)

into Eq. (A2):

$$\tau_R \frac{\partial q_\alpha}{\partial t} + q_\alpha = -\frac{1}{3} C_V v_g^2 \tau_R \frac{\partial T}{\partial x_\alpha} + \frac{1}{5} \tau_C \tau_R v_g^2 \left[ \frac{\partial^2 q_\alpha}{\partial x_\beta \partial x_\beta} + \frac{1}{3} \frac{\partial}{\partial x_\alpha} \left( \frac{\partial q_\beta}{\partial x_\beta} \right) \right]. \tag{A7}$$

The non-equilibrium phonon distribution corresponding to Eq. (A7) is derived

from:

$$\left( \frac{\partial f}{\partial t} + v_{g\alpha} \frac{\partial f}{\partial x_\alpha} \right) \bigg|_{f_4} = \frac{f_N^{eq}}{\tau_N} + \frac{f_R^{eq}}{\tau_R} - \frac{f}{\tau_C}. \tag{A8}$$

Substitution of Eq. (A4) into Eq. (A8) produces the result as:

$$f = f_R^{eq} + \frac{3}{C_V v_g^2} \frac{\partial f_R^{eq}}{\partial T} q_\alpha v_{g\alpha} + \frac{\tau_C}{C_V} \frac{\partial q_\alpha}{\partial x_\alpha} \frac{\partial f_R^{eq}}{\partial T} - \frac{3\tau_C}{C_V v_g^2} v_{g\alpha} v_{g\beta} \frac{\partial q_\alpha}{\partial x_\beta} \frac{\partial f_R^{eq}}{\partial T}. \tag{A9}$$

With the help of the nonequilibrium phonon distribution function Eq. (A9) and the

standard procedures in the kinetic theory [30, 35, 36], the boundary conditions for

temperature and heat flux are derived separately as:

$$T_s - T = -\frac{2}{3} \Lambda_R \left( \frac{\partial T}{\partial n} \right)_s, \tag{A10}$$

$$\mathbf{q}_s = \frac{8}{15} \Lambda_C \left( \frac{\partial \mathbf{q}}{\partial n} \right)_s, \tag{A11}$$

The diffuse phonon-surface scattering has been assumed in developing the boundary conditions.

The phonon hydrodynamic equation (9) has been demonstrated to be valid for nanoscale heat transport around ordinary temperatures at both a spatial and temporal Knudsen numbers smaller than 0.3, *i.e.* within the slip regime and early transitional regime [30]. Therefore, the present phonon hydrodynamic equation (A7) together with the non-equilibrium boundary conditions Eq. (A10) and Eq. (A11) is expected to predict overall accurate results for $Kn_C \leq 0.3$. Here we provide a validation of Eq. (A7) within the slip regime ($Kn_C \leq 0.1$), which covers the near-continuum regime ($Kn_C \leq 0.01$) where the lattice Boltzmann scheme developed in Section 2 is valid. Thus we can use the analytical solution of phonon hydrodynamic model as the benchmark of the lattice Boltzmann scheme in Section 3. The classical in-plane and cross-plane phonon transport through a thin film with a thickness $d$ shown in Figure 6 are taken into account. The material phonon properties are the same as those in Section 3.

For the in-plane phonon transport in Figure 6(a), we considered two cases of $Kn_N = 0.05$ and $Kn_N = 0.1$. For each case, we vary $Kn_R$ from the regime where the resistive scattering dominates over the normal scattering to the regime where the normal scattering dominates over the resistive scattering. The analytical solution of heat flux distribution based on the phonon hydrodynamic model has been obtained in Eq. (68), which is compared to the DOM solution of the phonon Boltzmann equation under Callaway's dual relaxation model [13] in Figure A1. The analytical solutions of both G-K heat transport equation with the heat flux slip boundary condition and Fourier's law are included for a systematic comparison. A very good agreement is obtained between the analytical results by phonon hydrodynamic equation and the direct DOM numerical solution in different transport regimes. When the normal scattering is dominant over the resistive scattering, phonon transport lies in the hydrodynamic regime where phonon Poiseuille flow with parabolic heat flux profile is obtained, as seen for the case $Kn_N = 0.05$ with $Kn_R = 5$, $Kn_R = 50$, $Kn_R = 500$ in Figure A1(a), and for the case $Kn_N = 0.1$ with $Kn_R = 10$, $Kn_R = 100$ in Figure A1(b). The result of phonon hydrodynamic equation becomes consistent with that of G-K heat transport equation which is derived in the limit of dominant normal scattering. With the decrease of $Kn_R$, the phonon Poiseuille flow is significantly destroyed by the increasing resistive scattering, which tends to introduce a uniform thermal transport resistance throughout the medium. A deviation of the result of phonon hydrodynamic model from that of G-K heat transport equation emerges as shown for the case $Kn_N$=0.05 with $Kn_R$=0.01, $Kn_R$=0.05 in Figure A1(a), and for the case $Kn_N$=0.1 with $Kn_R$=0.01, $Kn_R$=0.1 in Figure A1(b). In the limit of absolutely dominant resistive scattering over the normal scattering, the phonon hydrodynamic equation will be reduced to the classical Fourier's law. In all, the present phonon hydrodynamic model captures well the phonon transport behavior throughout the entire transition from the hydrodynamic regime to diffusive regime.

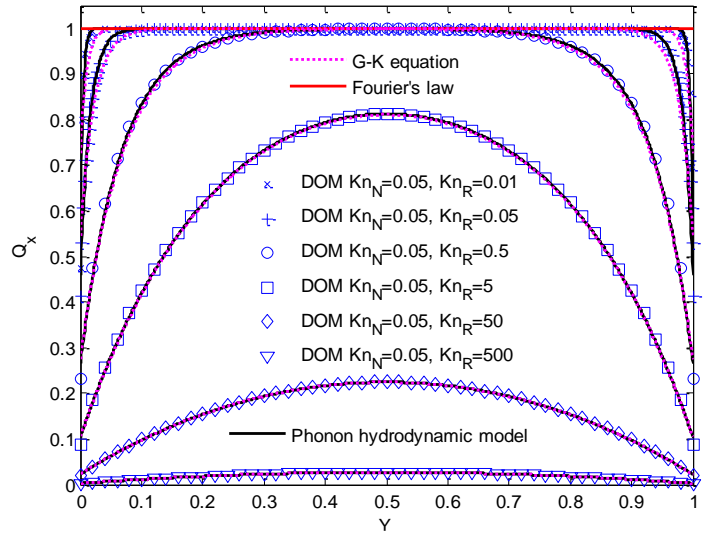

(a)

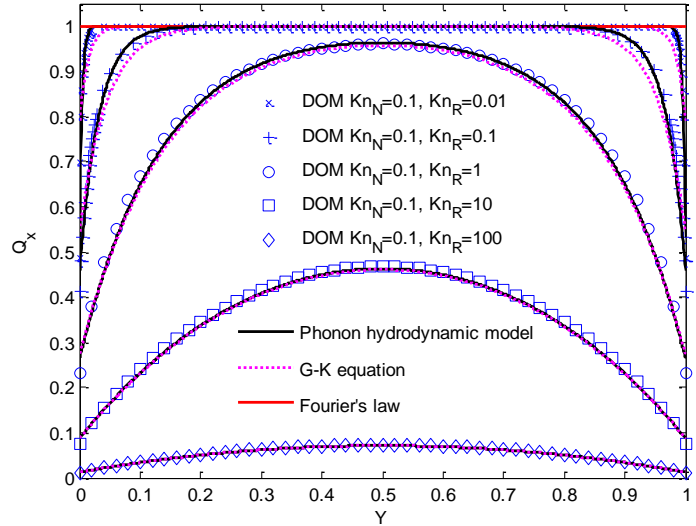

(b)

Figure A1. Non-dimensional heat flux distribution in in-plane phonon transport: (a) $Kn_N$=0.05, (b) $Kn_N$=0.1, the symbols represent the numerical solution of phonon Boltzmann equation under Callaway's dual relaxation model by discrete-ordinate-method (DOM) [13] with a grid of 3×2002 and angular resolution $N_\mu$×$N_\varphi$=16×16 after independence verification, whereas the solid black

lines, the solid red lines and the dashed lines represent respectively the analytical solutions of the present phonon hydrodynamic model, G-K heat transport equation and Fourier's law.

For the cross-plane phonon transport in Figure 6(b), we considered three cases of $Kn_N=0.001$, $Kn_N=0.01$ and $Kn_N=0.1$. For each case, we vary the resistive Knudsen number from 0.01 to 100. The analytical solutions of temperature distribution and heat flux based on the phonon hydrodynamic model have been obtained in Eq. (70) and Eq. (71), which agree well with the DOM solution of phonon Boltzmann equation under Callaway's dual relaxation model [13], as shown in Figure A2. In the limit of small resistive Knudsen number, the phonon hydrodynamic equation will be reduced to the classical Fourier's law.

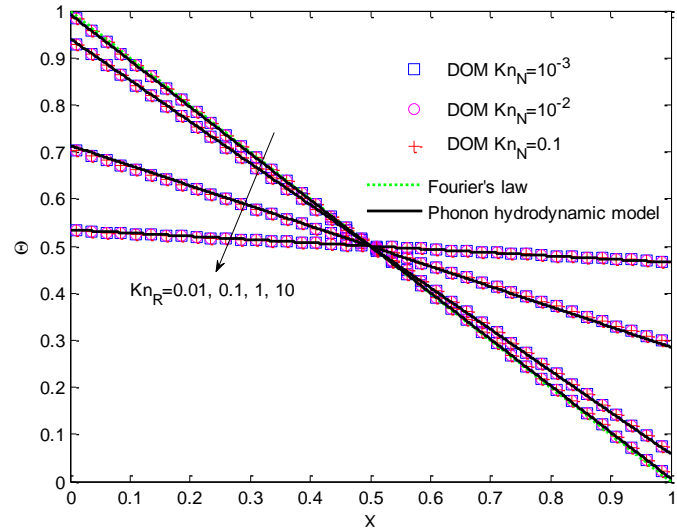

(a)

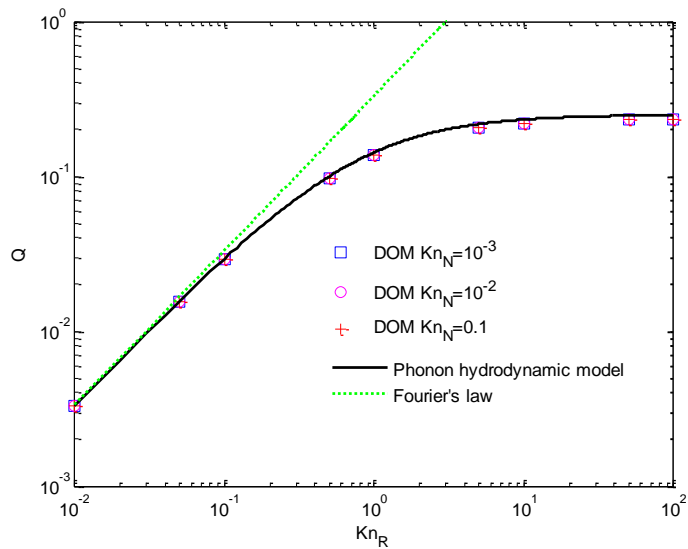

(b)

Figure A2. Non-dimensional temperature distribution and heat flux in cross-plane phonon transport:

(a) temperature distribution, (b) heat flux, the symbols represent the numerical solution of phonon Boltzmann equation under Callaway's dual relaxation model by discrete-ordinate-method (DOM) [13] with a grid of 201 and angular resolution $N_\mu$=16 after independence check, whereas the solid lines and dashed lines represent respectively the analytical solutions of the present phonon hydrodynamic model and the Fourier's law.

# Appendix B. Laplace transform solution of phonon hydrodynamic model for transient heat pulse propagation

The following Laplace transforms of the non-dimensional temperature and heat flux are introduced as:

$$\overline{\Theta}(X;s) = \int_0^\infty \Theta(X,t^*) \exp(-st^*) dt^*, \tag{B1}$$

$$\overline{Q}(X;s) = \int_0^\infty Q(X,t^*) \exp(-st^*) dt^*, \tag{B2}$$

where $s$ is a complex parameter. With the aid of initial condition Eq. (79), the Laplace transform of the partial time derivative of non-dimensional temperature and heat flux becomes:

$$\int_0^\infty \frac{\partial \Theta(X,t^*)}{\partial t^*} \exp(-st^*) dt^* = s\overline{\Theta}(X;s), \tag{B3}$$

$$\int_0^\infty \frac{\partial Q(X,t^*)}{\partial t^*} \exp(-st^*) dt^* = s\overline{Q}(X;s). \tag{B4}$$

Therefore, the non-dimensional governing equations (77) and (78) are reduced to:

$$s\overline{\Theta}(X;s) + Kn_N \frac{d\overline{Q}(X;s)}{dX} = 0, \tag{B5}$$

$$\left(\frac{Kn_R}{Kn_N}s + 1\right)\overline{Q}(X;s) = -\frac{1}{3}Kn_R \frac{d\overline{\Theta}(X;s)}{dX} + \frac{4}{15}Kn_C Kn_R \frac{d^2\overline{Q}(X;s)}{dX^2}. \tag{B6}$$

After a Laplace transform, the boundary condition Eq. (80) becomes:

$$\begin{cases} X=0, & \overline{Q} = \frac{1}{s}\left[1-\exp(-100s)\right] \\ X=1, & \overline{Q} = 0 \end{cases}. \tag{B7}$$

Through eliminating the temperature term in Eqs. (B5) and (B6), we acquire the governing equation for heat flux:

$$\frac{d^2\overline{Q}(X;s)}{dX^2} = A^2\overline{Q}(X;s), \tag{B8}$$

where the parameter A is introduced for short notation:

$$A \equiv \sqrt{\frac{\dfrac{Kn_R}{Kn_N}s+1}{\dfrac{1}{3}\dfrac{Kn_N Kn_R}{s}+\dfrac{4}{15}Kn_C Kn_R}} \quad .$$

(B9)

The general solution of Eq. (B8) is:

$$\overline{Q}(X;s)=C_1\exp(AX)+C_2\exp(-AX) \, .$$

(B10)

The coefficients $C_1$ and $C_2$ in Eq. (B10) are determined based on the boundary conditions Eq. (B7):

$$C_2=\frac{1}{s}\frac{1-\exp(-100s)}{1-\exp(-2A)} \, ,$$

(B11)

$$C_1=-C_2\exp(-2A) \, .$$

(B12)

The non-dimensional temperature distribution in the Laplace-transform domain is related to the non-dimensional heat flux distribution Eq. (B10) through Eq. (B5) as:

$$\overline{\Theta}(X;s)=-\frac{Kn_N}{s}\left[C_1 A\exp(AX)-C_2 A\exp(-AX)\right].$$

(B13)

The temporal evolutions of the temperature and heat flux distributions are resolved through an inverse Laplace transform based on the Riemann-sum approximation [6]:

$$\Theta\left(X,t^*\right)=\frac{\exp\left(\gamma t^*\right)}{t^*}\left[\frac{1}{2}\overline{\Theta}(X;\gamma)+\mathrm{Re}\sum_{n=1}^{\infty}\overline{\Theta}\left(X;\gamma+\frac{in\pi}{t^*}\right)(-1)^n\right],$$

(B14)

$$Q\left(X,t^*\right)\simeq\frac{\exp\left(\gamma t^*\right)}{t^*}\left[\frac{1}{2}\overline{Q}(X;\gamma)+\mathrm{Re}\sum_{n=1}^{\infty}\overline{Q}\left(X;\gamma+\frac{in\pi}{t^*}\right)(-1)^n\right],$$

(B15)

where $\gamma t^*=3$, and 'Re' denotes the real part of a complex variable, with '$i$' the imaginary index. The Riemann-sum approximation in Eq. (B14) and Eq. (B15) is actually a special case of a more general approximate formula for inverse Laplace transform [59]:

$$f\left(t\right)=\frac{\exp(\gamma t)}{T_p}\left[\frac{1}{2}F\left(\gamma\right)+\mathrm{Re}\sum_{n=1}^{\infty}F\left(\gamma+\frac{in\pi}{T_p}\right)\exp\left(\frac{in\pi t}{T_p}\right)\right],$$

(B16)

where $f(t)$ is the inverse Laplace transform of $F(\gamma)$, and $2T_p$ is the period of a periodic function $g_0(t)$. The basic idea of the derivation of Eq. (B16) is to obtain the complex Fourier series for the function $g_0(t)$ and equal to $f(t)e^{-\gamma t}$ on the interval $(0, 2T_p)$ [59]. The error of the approximation in Eq. (B16) has been shown to be as small as desired within $0 < t < 2T_p$ by choosing an adequate $\gamma$ [59]. Therefore, Eq. (B14) and Eq. (B15) is obtained from Eq. (B16) by adopting $T_p = t$. Note the recommended value of $\gamma t^* \cong 4.7$  [6] yields nearly the same results except very small oscillations near the heat pulse boundary at long time. Thus we choose a slightly different value of $\gamma t^* = 3$, which produces less oscillatory results.